\documentclass[%
 reprint,
 amsmath,amssymb,
 aps,
]{revtex4-2}

\usepackage{graphicx}
\usepackage{dcolumn}
\usepackage{bm}
\usepackage{xcolor}
\usepackage{ulem}
\usepackage{physics} 
\usepackage{float}
\usepackage{soul} 

\usepackage{hyperref} 

\UseRawInputEncoding

\begin{document}
\preprint{APS/123-QED}

\title{Exchange anisotropies in microwave-driven singlet-triplet qubits} 

\author{Jaime Saez-Mollejo$^1$, Daniel Jirovec$^{1,2}$, Yona Schell$^1$, Josip Kukucka$^1$, Stefano Calcaterra$^3$, Daniel Chrastina$^3$, Giovanni Isella$^3$, Maximilian Rimbach-Russ$^2$, Stefano Bosco$^2$ and Georgios Katsaros$^1$}
\affiliation{$^1$ Institute of Science and Technology Austria, Klosterneuburg, Austria.}
\affiliation{$^2$ QuTech, Delft University of Technology, Delft, The Netherlands.}
\affiliation{$^3$ Laboratory for Epitaxial Nanostructures on Silicon and Spintronics, Physics
Department, Politecnico di Milano, Como, Italy.}

\date{\today}

\begin{abstract}
Hole spin qubits are rapidly emerging as the workhorse of semiconducting quantum processors because of their large spin-orbit interaction, enabling fast all-electric operations at low power. However, spin-orbit interaction also causes non-uniformities in devices, resulting in locally varying qubit energies and site-dependent anisotropies. While these anisotropies can be used to drive single-spins, if not properly harnessed, they can hinder the path toward large-scale quantum processors. Here, we report on microwave-driven singlet-triplet qubits in planar germanium and use them to investigate the anisotropy of two spins in a double quantum dot. We show two distinct operating regimes depending on the magnetic field direction. For in-plane fields, the two spins are largely anisotropic, and electrically tunable, which enables to measure all the available transitions; coherence times exceeding 3 $\mu$s are extracted.
For out-of-plane fields, they have an isotropic response but preserve the substantial energy difference required to address the singlet-triplet qubit. Even in this field direction, where the qubit lifetime is strongly affected by nuclear spins, we find 400 ns coherence times. Our work adds a valuable tool to investigate and harness the anisotropy of spin qubits and can be implemented in any large-scale NxN device, facilitating the path towards scalable quantum processors. 
\end{abstract}

\maketitle


Semiconductor spin qubits are promising candidates for the realization of compact integrated quantum circuits \cite{vandersypen_interfacing_2017, veldhorst_silicon_2017}. Unlike electron spins, hole spin qubits have a strong intrinsic spin-orbit interaction (SOI) that permits fast electrical driving without the need for on-chip micromagnets, simplifying scalability requirements. Four-qubit processors have already been demonstrated~\cite{hendrickx_four-qubit_2021, zhang_universal_2024} and a 10-qubit device presented~\cite{wang_operating_2024}. However, the SOI also introduces a sensitivity of the g-tensor to variations in the strength of electric fields, the confinement potential, and the strain in the heterostructure~\cite{bosco_squeezed_2021, abadillo-uriel_hole-spin_2023, martinez_hole_2022, sarkar_electrical_2023, wang_electrical_2024}. This sensitivity results in site-dependent qubit properties, with each quantum dot having a unique g-tensor, consequently tilting the quantization axis of each qubit with respect to its neighbours. While this property has been recently used for driving coherent spin rotations and demonstrating hopping-based universal quantum logic ~\cite{van_riggelen-doelman_coherent_2024, wang_operating_2024}, it brings several challenges as the g-tensor variability can affect the Pauli Spin Blockade read-out, the energy spectrum, the efficiency of the driving mechanism and the noise susceptibility~\cite{sen_classification_2023, qvist_probing_2022}. 

To address this challenge, we take advantage of the large g-factor anisotropy of heavy-holes states~\cite{watzinger_heavy-hole_2016}. In particular, for planar (001) Ge/SiGe heterostructures, the effective g-factor for out-of-plane fields is more than ten times larger than the one for in-plane fields~\cite{jirovec_dynamics_2022, hendrickx_sweet-spot_2024}. Therefore, for an out-of-plane field, the crystallographic and qubit axes are very similar, minimising the g-tensor variability among different qubits. However, so far all microwave-driven and most baseband-controlled qubits in natural planar Ge have been realized for in-plane magnetic fields in order to minimize the hyperfine interaction to randomly distributed nuclear spins, as it is of Ising-type, pointing in the direction of strongest confinement~\cite{fischer_spin_2008, fischer_hybridization_2010, bosco_fully_2021, hendrickx_sweet-spot_2024}. 

Here, we make use of a double quantum dot device formed in a Ge/SiGe heterostructure to characterize the exchange anisotropy, a key quantity for future scaling of semiconductor spin qubit quantum processors. We develop a simple protocol which allows to systematically characterize all the relevant spectroscopic features of two-hole systems at fixed magnetic field directions, including the angle between the quantization axes. For in-plane fields, we investigate microwave-driven singlet-triplet transitions --including the elusive $T_-$ to $T_+$ transition-- and we show that such transitions are driven via the modulation of the exchange interaction and are just possible because of g-tensor dissimilarities. We further demonstrate that the quantization axes misalignment of the two spins can be electrically tuned. For out-of-plane fields, we demonstrate spin axes alignment and realize a $S-T_{0}$ microwave-driven qubit, which already for small fields has a large energy splitting and small transition matrix elements to the leakage states $T_-$ and $T_+$~\cite{mutter_all-electrical_2021}. This solves the problem of the non-orthogonal rotation axes, which have been limiting their implementation. 

\section*{Main}
\begin{figure*}[t]
\includegraphics[width=\textwidth]{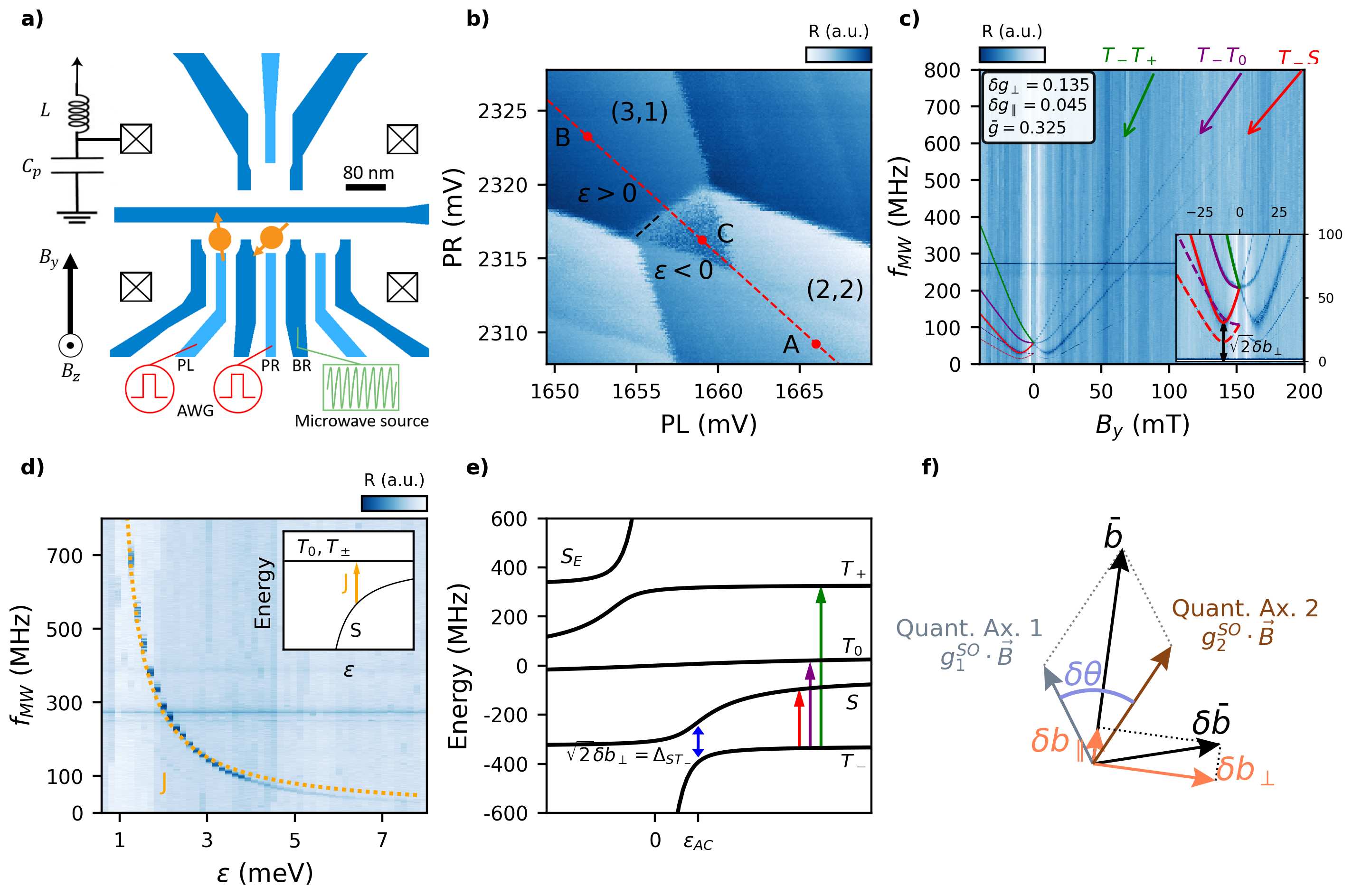}
\caption{(a) Device schematics showing the gate layout. The two bottom rightmost gates are kept at 0 V. Fast pulses are applied to gates PL and PR to change the charge occupation of the DQD, while microwave bursts are applied to the right barrier BR to induce spin transitions. The in-plane magnetic field $B_y$ is applied perpendicular to the axis of the DQD. (b) Stability diagram of the investigated transition. The dark blue triangle in the (2,2) regime corresponds to the Pauli Spin Blockade region. This colour code is maintained throughout the paper, where dark corresponds to blocked states and light to unblocked. The red dashed line indicates the device's operational detuning line. (c, d) Charge sensor amplitude as a function of $B_y$, $\varepsilon$ and $f_{MW}$. The increase in the signal corresponds to a higher triplet return probability, reflecting a change in the system's level population.  In (c) the dependence is shown at detuning $\varepsilon = 4.6$ meV and in (d)  at $B_y=0$ mT. The solid lines are the transition frequencies obtained when using the extracted parameters $\bar{g}$, $\delta g_\parallel$ and $\delta g_\perp$.  The inset in c) shows a low-field high-resolution zoom-in. In both cases, the duration of the microwave burst is 10 $\mu s$ and the readout time 1 $\mu s$.  From the transition frequency in (d) we can extract the exchange interaction versus detuning by converting the voltage amplitude of the pulse into energy with the lever arms (see Fig.~\ref{SFig:SuppLeverArms}(a)). By fitting (orange dotted line) to the expression $J=-\varepsilon /2 + \sqrt{\varepsilon^2/4 + 8t_c^2}$ ~\cite{jirovec_singlet-triplet_2021}, we obtain a tunnel coupling of $t_c/h=6.6\pm 0.1$ GHz. (e) Energy diagram of the four investigated states versus detuning. The anticrossing between the $\ket{S}$ and $\ket{{T}_-}$ states occurs at $\varepsilon_{AC}$, having two consequences: the ground state changes and defines the minimum energy difference between the two lowest energy states, $\Delta_{ST_-}$. (f) Schematic explaining the quantization axes (grey and brown), the addition and difference of the Zeeman vectors (black) as well as the projections $\delta b_{\parallel}$ and $\delta b_{\perp}$ (orange), which enable spin driving via exchange modulation.} 
\label{Fig:Fig1}
\end{figure*}

The schematics of the device used in the experiment is shown in Fig.~\ref{Fig:Fig1}(a).  Electrostatic gates deplete the two-dimensional hole gas beneath, creating a double quantum dot (DQD) on the bottom part of the device to be operated as a qubit. A single quantum dot on the top part is used as a charge sensor whose impedance is measured via radio-frequency reflectometry ~\cite{schoelkopf_radio-frequency_1998, vigneau_probing_2023}. 

The double quantum dot is operated in the (3,1)-(2,2) charge transition, where ($n_L$, $n_R$) correspond to the occupation of the left and right dots, respectively. This transition is equivalent to the (1,1)-(0,2) transition used for spin-selective readout via Pauli Spin Blockade, however, no spin dynamics were observed in the latter transition for unclear reasons \footnote{Pauli Spin Blockade was observed in the (1,1)-(0,2) charge transition; however, no time-resolved singlet-triplet oscillations were detected for different voltage configurations of the middle and outer barriers.}. In Fig.~\ref{Fig:Fig1}(b), a stability diagram illustrates the charge transition of interest, featuring the characteristic metastable triangle indicative of Pauli Spin Blockade~\cite{ono_current_2002}. This measurement was taken while pulsing the gates PR and PL following the path depicted by the points A, B and C along the detuning line, $\varepsilon$ (red dashed line). The system is initialized in a singlet state S(2,2) represented by A, and then a fast pulse is applied to move to B, where the spins are separated. At the measurement point C, the triplet states are blocked, forming the characteristic metastable triangle. We adopt the convention of $\varepsilon > 0$ for the (3,1) occupation and $\varepsilon < 0$ for the (2,2) state. 

In order to drive the spin transitions, the DC voltages are fixed at C and the system is first pulsed to the (2,2) singlet state (A), followed by a pulse with a 600 ns ramp into the (3,1) state, initializing in the ground state (B). Subsequently, a microwave burst is applied, inducing a spin-flip transition if the energy difference between the system's eigenstates, $\Delta E$, matches the burst energy $h f_{MW} = \Delta E$, where h is Planck's constant and $f_{MW}$ the frequency of the burst.  Finally, the system is brought back to point C inside the Pauli Spin Blockade triangle and the spin readout is performed.

\section*{In-plane magnetic field}
We start by mapping out the amplitude of the signal at the Pauli Spin Blockade point as a function of the in-plane magnetic field $B_y$ and frequency of the microwave burst as seen in Fig.~\ref{Fig:Fig1}(c). We observe the appearance of three spin transitions, in contrast to the typically observed two lines ~\cite{nadj-perge_spinorbit_2010, camenzind_hole_2022, hendrickx_fast_2020}. Those transitions are due to spin flips of charges localized in the two QDs. The faint line at the lowest frequency has half the frequency of the line indicated by the red arrow. This subharmonic transition is a consequence of a non-linear driving mechanism (see Fig.~\ref{SFig:SuppHarmonics}) ~\cite{stehlik_extreme_2014,scarlino_second-harmonic_2015}. 

 As the magnetic field is reduced below 30 mT, the 3 lines bend and asymptotically merge at zero magnetic field. Similar features have been observed for InSb and SiMOS devices at low magnetic fields, and the observation was attributed to exchange interaction ~\cite{nadj-perge_spectroscopy_2012, ono_hole_2017}. As a first step, and to characterize the strength of the exchange interaction, $J$, we perform the spectroscopy as a function of detuning at $B=0$ mT as shown in Fig.~\ref{Fig:Fig1}(d). When the magnetic field is zero, the three triplets are degenerate and the only relevant energy scale is $J$. Exchange can be tuned from 800 MHz at low detuning to 25 MHz at the highest detuning we can experimentally reach. 

To understand the bending of the transition frequencies as a function of applied magnetic field and their physical implications, we introduce the following Hamiltonian suitable for hole spin qubits in the (1,1) charge occupation expressed in the $\{|S \rangle$, $|T_+ \rangle$, $|T_- \rangle$, $|T_0\rangle \}$ basis. Its eigenenergies and eigenstates are shown in Fig.~\ref{Fig:Fig1}(e) ~\cite{geyer_anisotropic_2024, ungerer_coherence_2024, stepanenko_singlet-triplet_2012} (see Methods for details). 

\begin{align}
    H=&-J\ket{S}\bra{S}+\bar{b}(\ket{T_+}\bra{T_+}-\ket{T_-}\bra{T_-})\nonumber\\
    +&\sin{\theta}\left[\frac{\delta b_\perp}{\sqrt{2}}(\ket{S}\bra{T_-}-\ket{S}\bra{T_+})+\delta b_{\parallel}\ket{S}\bra{T_0} \right] + \text{h.c.}  
    \label{eq:static_H}
\end{align}
where $\theta=- \text{arctan2}(\varepsilon,\sqrt{2}t_c)/2$ is the mixing angle, $J=-\varepsilon /2 + \sqrt{\varepsilon^2/4 + 8t_c^2}$ is the exchange interaction, with $t_c$ the tunnel coupling. $\boldsymbol{b}$ ($\delta \boldsymbol{b}$) is the addition (difference) of the Zeeman vectors $\mu_B \underline{g}_{i}^{\text{so}}\boldsymbol{B}/2$ of the two dots ($i=1,2$), where $\mu_B$ is the Bohr magneton and $\boldsymbol{B}$ is the magnetic field. The parallel and perpendicular projections of $\delta\boldsymbol{b}$ on $\boldsymbol{b}$ are denoted $\delta b_{\parallel}$ and $\delta b_{\perp}$, respectively (see Fig.~\ref{Fig:Fig1}(f)). For convenience, we introduce the corresponding dimensionless g-factors as $\bar{b}=\mu_B |(\underline{g}_1^{\text{so}} +\underline{g}_2^{\text{so}}) \boldsymbol{B}|/2 =\mu_B \bar{g} |\boldsymbol{B}|$, $\delta b_{\parallel}= \mu_B \delta g_\parallel |\boldsymbol{B}|$ and $\delta b_{\perp}=\mu_B \delta g_\perp |\boldsymbol{B}|$. It is important to note that $\bar{g}$, $\delta g_\parallel$ and $\delta g_\perp$ are effective g-factors, which also incorporate the spin-flip tunnelling effects ~\cite{stepanenko_singlet-triplet_2012, geyer_anisotropic_2024, ungerer_coherence_2024}. 

We determine all the parameters of the Hamiltonian which describe Fig.~\ref{Fig:Fig1}(c), using the following systematic protocol. First, we extract $J$ from the transition frequency at zero magnetic field for a given $\varepsilon$. From the $S - T_-$ anticrossing at magnetic field $B^*$, we obtain the value of $\bar{g}$ as $\bar{g}=J/(\mu_B B^*)$ and from the amplitude of the anticrossing we obtain $\delta g_\perp$ as it is given by $\Delta_{ST}=\sqrt{2}\mu_B \delta g_\perp B^*$. Finally, $\delta g_\parallel$ is extracted from the slope of the transitions at high fields (see Methods). From this procedure, the values obtained for in-plane magnetic fields are $\bar{g}=0.325$, $\delta g_\parallel=0.045$ and $\delta g_\perp =0.135$.  

As $\delta g_\perp \neq 0$, we conclude that the quantization axes of both quantum dots are not aligned, in analogy to similar experiments for hole spins confined in quantum wells operated in accumulation mode \cite{van_riggelen-doelman_coherent_2024, wang_operating_2024, zhang_universal_2024, hendrickx_sweet-spot_2024}.

We emphasize that an equivalent description appropriate for large B field comprises two aligned spins coupled by an anisotropic exchange interaction matrix $J R_y(-\delta \theta)$. The rotation angle is given by $\delta\theta=\text{arctan}\left[\delta g_\perp/(\bar{g}+ \delta g_\parallel)\right]+\text{arctan}\left[\delta g_\perp/(\bar{g}- \delta g_\parallel)\right]$ (see Methods).  This angle is the misalignement between both quantization axes (see Fig. \ref{Fig:Fig1}(f) and Fig. \ref{SFig:SuppGeo}). From the above experimentally extracted values we find $\delta\theta=45^\circ$, highlighting the large degree of exchange anisotropy in our system. We note that large anisotropy is expected to be useful for resonant two-qubit gates in Loss-DiVincenzo qubits and for leakage protected two-qubit gates in ST qubits. Therefore, the ability to electrically tune this anisotropy could serve as a crucial control knob for qubit operation.

\begin{figure}[t]
\includegraphics[width=\linewidth]{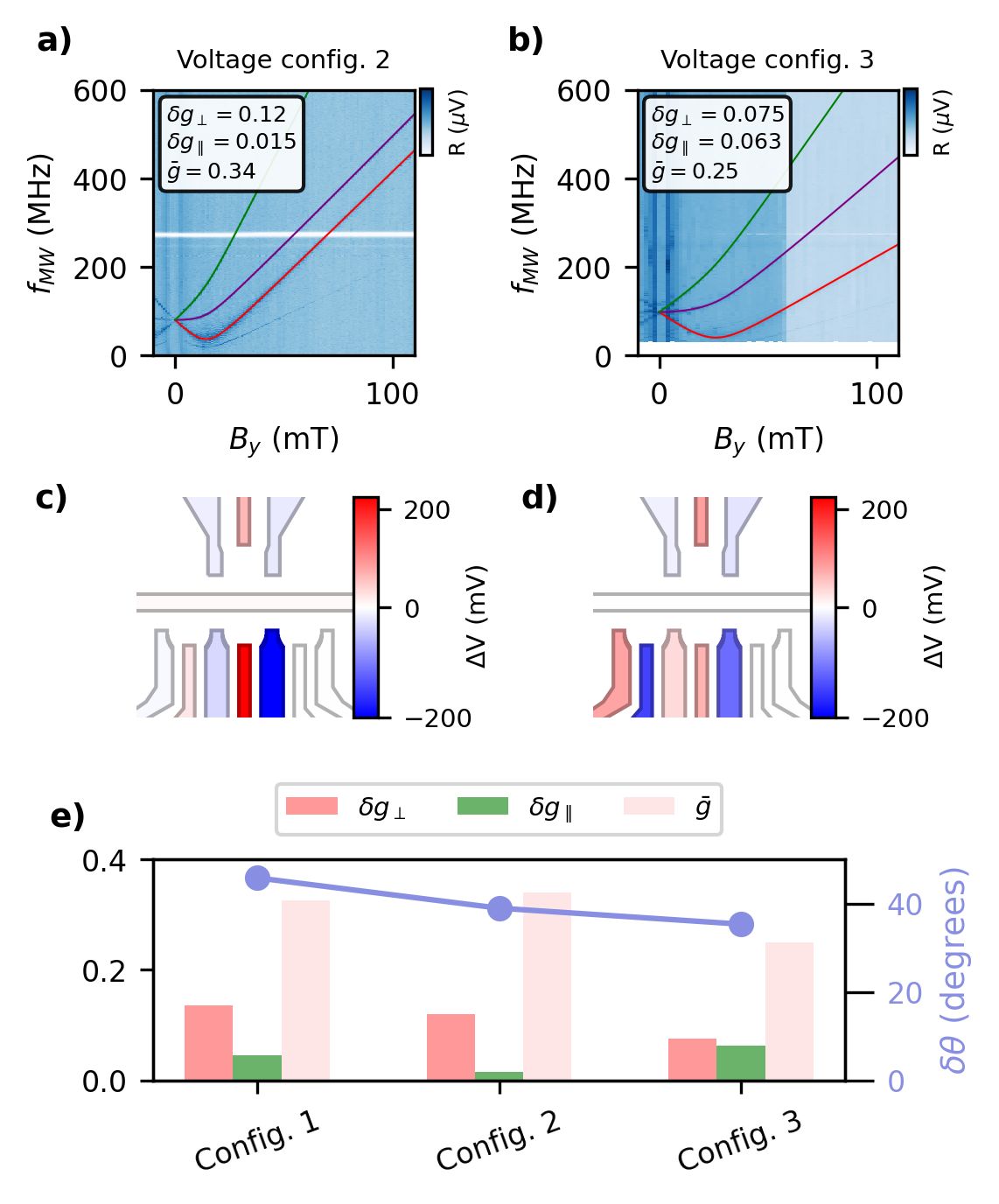}
\caption{Electrical tunability of $\bar{g}$, $\delta g_\perp$, $\delta g_\parallel$ and the quantization axes tilt. (a, b) Magnetic field spectroscopy for two different voltage configurations of the electrostatic gates. The colored solid lines are the transitions corresponding to the parameters written in the legend. Panels (c, d) show the voltage differences of each configuration with respect to the voltage configuration 1 (Fig. \ref{Fig:Fig1} (c)). (c) Extracted parameters for each electrostatic configuration, including the misalignment $\delta \theta$, demonstrating electrical tunability.}
\label{Fig:Fig2_new}
\end{figure}

Previous studies have independently demonstrated the electrical tunability of the g-tensor \cite{crippa_electrical_2018, liles_electrical_2021},  and the site-dependent properties between neighboring dots for a fixed electrostatic potential \cite{van_riggelen-doelman_coherent_2024, hendrickx_sweet-spot_2024}. To investigate the electrical tunability of the anisotropy of the dots, we tuned the DQD into two additional voltage configurations and repeated the measurement shown Fig.~\ref{Fig:Fig1}(c). Figure \ref{Fig:Fig2_new} (a, b) shows the magnetic field spectroscopy at those two different electrostatic potentials; the  $T_- - S$, $T_- - T_0$ and $T_- - T_+$ are again observed. As a consequence of the change in the electrostatic configuration (Figs. \ref{Fig:Fig2_new} (c, d)), the slopes of the transitions are different, implying that the tilt between the quantization axes changes. In voltage configuration 2 (Fig. \ref{Fig:Fig2_new} (a)), the transitions $T_- - S$ and $T_- - T_0$ are almost parallel, indicating that the Zeeman energy difference $\delta b_\parallel$ is small and that the $S-T_0$ splitting is dominated by exchange. On the other hand, the slopes in voltage configuration 3, (Fig. \ref{Fig:Fig2_new} (b)) are smaller and dissimilar. Figure \ref{Fig:Fig2_new}(e) compares the extracted parameters for the different voltage configurations. By calculating the relative tilt, we show the tunability of the misalignment by 10 degrees with gate voltage. This finding illustrates not only the utility of this protocol in quickly characterizing the site-dependent anisotropies, without extracting each g-tensor, but also the ability to control these anisotropies electrically.

\section*{Driving mechanism}
\begin{figure*}[t]
\includegraphics[width=\textwidth]{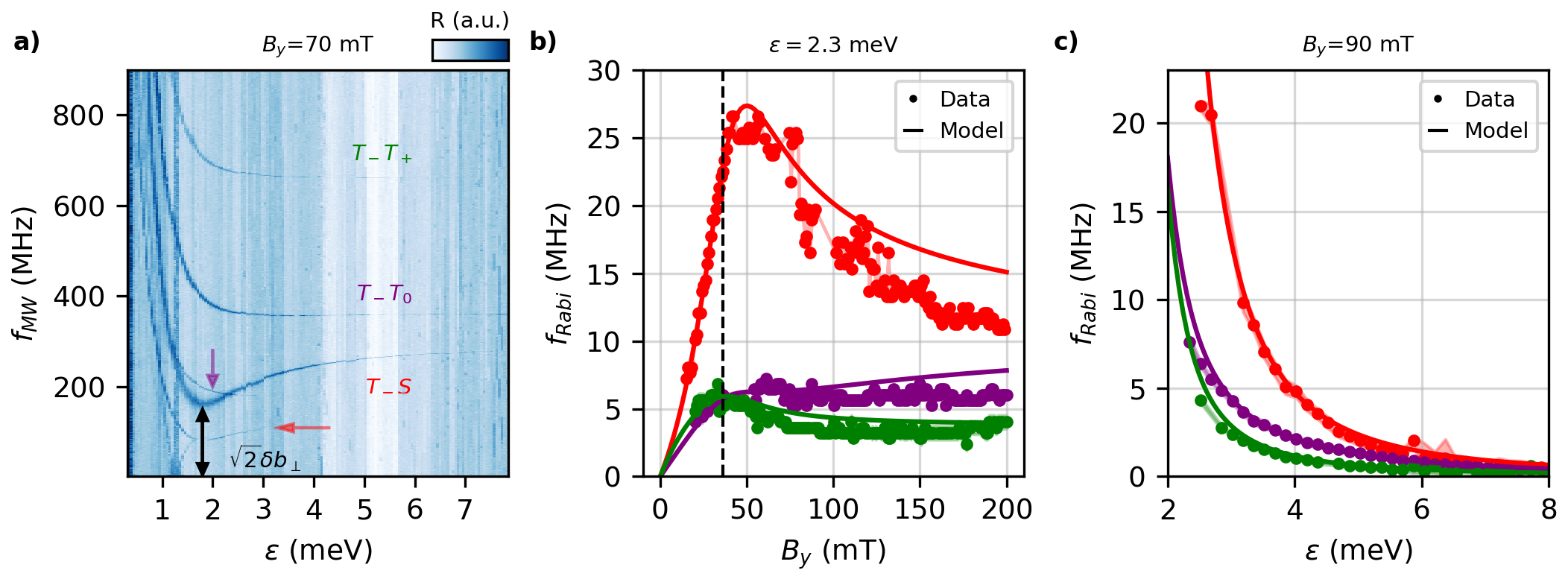}
\caption{(a)  Plot showing the transition frequency $f_{MW}$ versus detuning at an in-plane magnetic field of 70mT. The anticrossing (black arrow) occurs at $\varepsilon_{AC}=1.8$ meV and has an amplitude of $\Delta_{ST_-}= 175$ MHz. We note that after the avoided crossing, the green transition corresponds to a double spin-flip from ${T}_-\rightarrow  {T}_+$,  which usually is not observed in spin qubit systems. At low $\varepsilon$, we observe subharmonic transitions from ${T}_-- {S}$ and ${T}_--  {T}_0$ indicated by the red and purple arrows, respectively. This indicates that the non-linearities of the driving mechanism are non-negligible. (b)-(c) Rabi frequency versus magnetic field and detuning, respectively, following the same colour code as in Fig.~\ref{Fig:Fig1}(e). The solid lines in (b), (c) represent the exact solution for the Rabi frequencies with the parameters extracted from the Hamiltonian. The analytical expressions in the limits $J\ll|\delta\textbf{b}|$ and $J\gg|\delta\textbf{b}|$ can be found in the Supplementary. In (b) and for higher fields, we note a discrepancy between the experimental data and the numerical solution as the field increases, which can be at least partially attributed to the larger cable attenuation, i.e. smaller power arriving at the device, at higher frequencies.}
\label{Fig:Fig2}
\end{figure*}

After the characterization of the spectroscopic features, we return to the voltage configuration 1 and focus on extracting the mechanism which permits to drive these transitions. We study the dependence of the spin transitions as a function of detuning at a fixed magnetic field in Fig.~\ref{Fig:Fig2}(a). As the system is initialized in the ground state, the three observed lines correspond to the transitions indicated by the three arrows depicted in Fig.~\ref{Fig:Fig1}(e). From the $T_--S$ transition, we can experimentally determine the position, $\varepsilon_{AC}=1.8$ meV, and size, $\Delta_{ST_-}= 175$ MHz, of the $S-T_-$ anticrossing. As this point sets the turnover of the ground state, we identify two regimes. When $\varepsilon<\varepsilon_{AC}$, the ground state is a singlet state and we drive the transitions ${S}\rightarrow ({T}_-, {T}_0, {T}_+)$; when $\varepsilon>\varepsilon_{AC}$ the transitions are  ${T}_-\rightarrow ({S}, {T}_0, {T}_+)$. When $\varepsilon<\varepsilon_{AC}$, the Larmor frequency and the linewidth of the three transitions rapidly increases when reducing $\varepsilon$. When $\varepsilon>\varepsilon_{AC}$, the shifts in Larmor frequencies for the three transitions exhibit a diminishing trend due to the reduction of $J$ with $\varepsilon$. We also observe that the line-widths of all the transitions also become smaller with larger detuning (smaller exchange). 

We next study the evolution of the Rabi frequencies as a function of detuning and magnetic field for each transition.
Coherent Rabi oscillations between the ground state and the three excited states at $\varepsilon=2.3$ meV are measured and shown in Figs.~\ref{SFig:Supp_RabiChevrons} and~\ref{SFig:SupRabis25mV} allowing us to estimate the gate fidelities (see~\ref{fig:EstimatedFidelity}). From the fast Fourier transform of those measurements, we extract the Rabi frequencies versus magnetic field, which are shown in Fig.~\ref{Fig:Fig2}(b). While the transitions to the ${T}_0$ and the ${T}_+$ states do not show any strong magnetic field dependence above 60 mT, this is strikingly different for the $T_--S$ transition, which exhibits a non-monotonic behaviour with the highest Rabi frequency observed close to the position of the anticrossing, indicated by the dashed black line in Fig.~\ref{Fig:Fig2}(b). This behaviour contrasts with what is typically observed in systems with strong SOI, where the Rabi frequency scales linearly with the magnetic field ~\cite{nowack_coherent_2007, froning_ultrafast_2021, crippa_electrical_2018, hendrickx_sweet-spot_2024}. By fixing the magnetic field to $B_y=90$ mT and plotting the Rabi frequency versus detuning, we observe that now all transitions show a decreasing Rabi frequency as the exchange interaction is reduced.  This last observation indicates that exchange plays a vital role in the driving mechanism.

This exchange driving originates from the coupling of the ac-electric field applied on the BR gate, which predominantly affects the detuning of the DQD. The change in detuning produced by BR is $\delta \varepsilon_{ac}(t)=  \alpha_{BR} \Delta V  \cos(2\pi f_{MW} t)$, where $\Delta V$ is the voltage amplitude of the microwave burst arriving at the sample and $\alpha_{BR}$ the lever arm of BR on the right QD (see Fig.~\ref{SFig:SuppLeverArms}). As a consequence, a time-dependent modulation of exchange shifts the energy of the singlet state with respect to the triplets, as represented by the following linearized Hamiltonian in the ST basis: $H_\text{driving}=\delta\varepsilon(t)\cos^2(\theta)\ket{S}\bra{S}$ (see Methods for more details).

As the microwaves only couples to the system by modulating the frequency of the singlet state, only states that are hybridized with the singlet states are affected by the drive. By looking at the Hamiltonian, we realize that the mixing between states depend on $\delta b_{\parallel}$ and $\delta b_{\perp}$, therefore the more dissimilar the quantum dots are, the stronger the hybridization and consequentially the faster the Rabi frequency. Due to the large anisotropy of spins in our system, we observe transitions between all the states. 

The Rabi frequencies are estimated from the transition amplitudes $\bra{f}H_\text{driving}\ket{i}$ between initial state $i$ and final state $f$ using the spectroscopy parameters extracted from Fig.~\ref{Fig:Fig1}(c). This model shows excellent agreement with the experimental Rabi frequencies for the three transitions (see Fig.~\ref{SFig:SupSims45mVand20mT} for additional data), demonstrating that the primary driving mechanism is the modulation of exchange instead of the conventional driving methods.

\section*{Out-of-plane magnetic field}
\begin{figure*}[t]
\includegraphics[width=\linewidth]{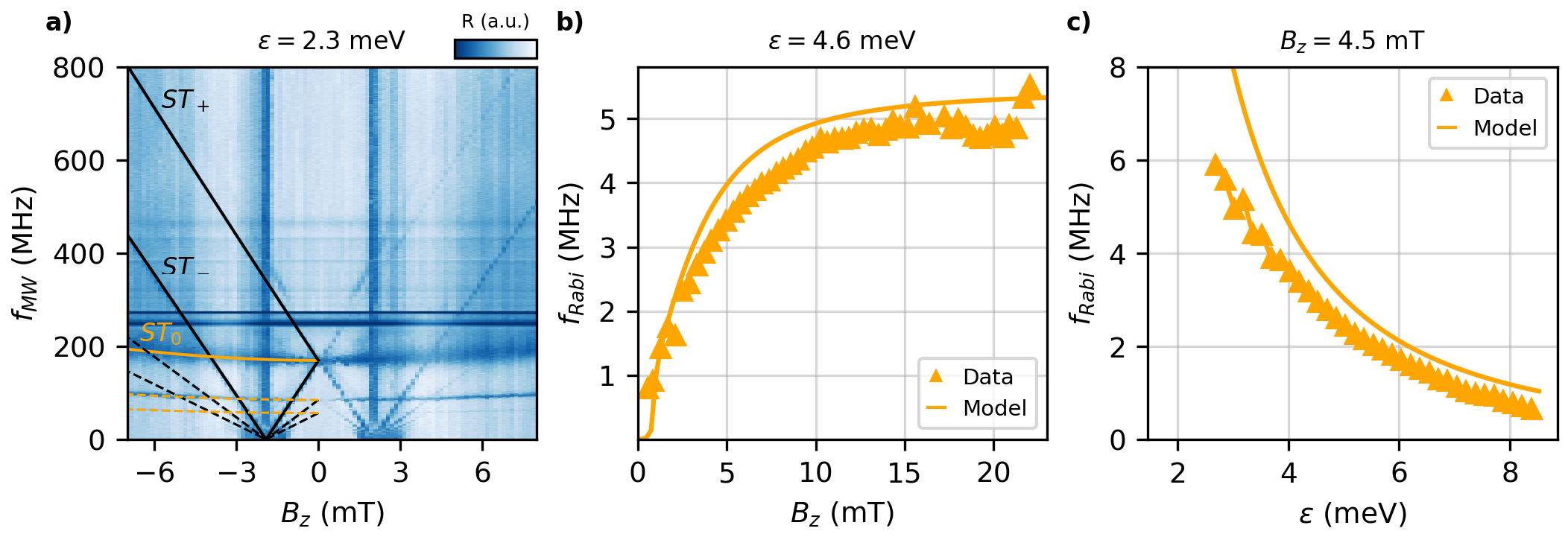}
\caption{(a) Out-of-plane magnetic field spectroscopy at $\varepsilon=2.3$ meV. The solid lines show the spin transitions from the singlet state to the triplets. The dashed lines correspond to the subharmonic transitions with half and a third of the frequency of the transitions. (b) and (c) show the measured Rabi frequencies as a function of $B_z$ and $\varepsilon$, respectively. The solid lines are the predicted $f_\text{Rabi}$ considering only exchange modulation. The error bars representing the standard deviation are smaller than the markers.}
\label{Fig:Fig3}
\end{figure*}

We now spectroscopically investigate the out-of-plane magnetic field direction, Fig.~\ref{Fig:Fig3}(a), in which the two quantization axes should be almost co-linear~\cite{hendrickx_sweet-spot_2024}. Similar to the previous case, we observe three lines and several subharmonic transitions, indicated by solid and dashed lines, respectively. However, in this case, the line with the lowest frequency does not show a clear anticrossing as in the in-plane case, with $\sqrt{2}\delta b_\perp$ being smaller than 30 MHz. With such a small $\delta b_\perp$ value, the system is always initialized in the singlet state for all implemented ramps, and therefore, the observed transitions are ${S}\rightarrow ({T}_-, {T}_0, {T}_+)$. By applying the same protocol as in the in-plane case, we obtain  $\bar{g}=6.6$, $\delta g_\parallel=0.5$ and $\delta g_\perp$ smaller than 0.3  (see Fig.~\ref{SFig:SupOutPlane}(c) for more details). This small value of $\delta g_\perp$ compared to $\bar{g}$ indicates that both spins are almost co-linear, however the finite value of $\delta g_\parallel$ provides the necessary Zeeman energy difference to address the $S-T_0$ qubit.

We focus on the ${S}-{T}_0$ transition, which has been investigated in both Si and Ge without the use of a microwave drive~\cite{jirovec_singlet-triplet_2021, liles_singlet-triplet_2023}. Indeed, coherent Rabi oscillations, whose frequency depends on the detuning are observed (see Fig.~\ref{SFig:SuppST0}). Figures \ref{Fig:Fig3}(b, c) summarize the dependence of $f_\text{Rabi}$ as a function of magnetic field and detuning of the $S-T_0$ qubit. The solid lines show the expected Rabi frequencies by considering only an exchange modulation driving as in the in-plane case. A very good agreement is also observed for the out-of-plane magnetic field direction. We note that the other two transitions only show a change in population but no coherent oscillations. We attribute this to the fact that $\delta b_{\perp}$ is small, leading to Rabi frequencies much smaller than the decoherence rate.

\section*{Comparison}
\begin{figure}[t]
\centering
\includegraphics[width=\linewidth]{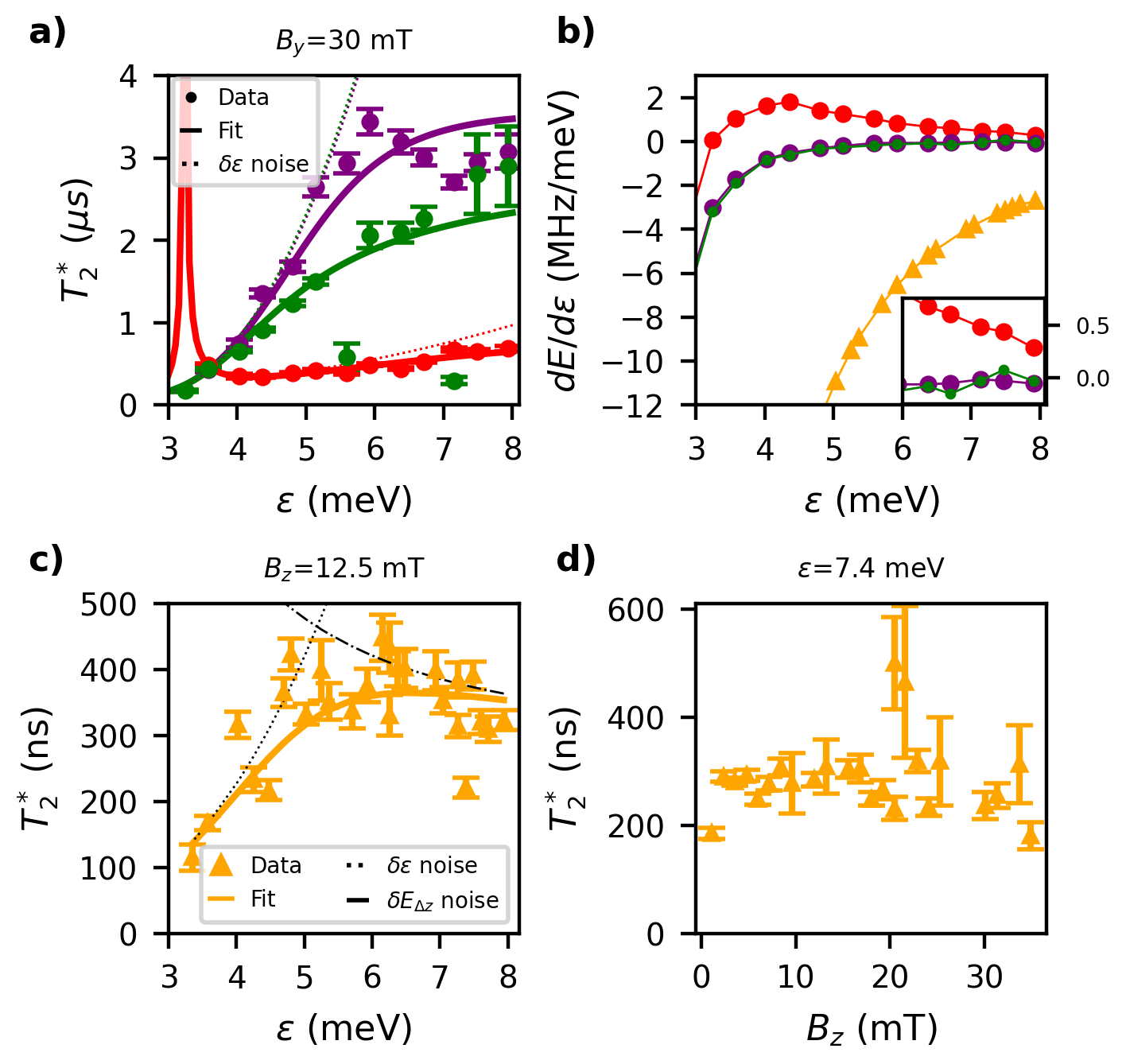}
\caption{(a) Inhomogeneous dephasing time for the three transitions at in-plane field as a function of detuning.  The colour code of each transition remains the same as previously described. Solid lines represent the fitted $T_2^\ast$ values for each transition, while the dashed lines indicate the dephasing time expected if only the detuning noise contribution is considered.  The purple and green dashed lines overlap and the peak predicted by the red fit (corresponding to the anticrossing) is not observed experimentally. Each measurement for extracting $T_2^\ast$ has been integrated for 20 minutes. The fitting parameters can be found in the Supplementary.  (b) Detuning susceptibility extracted from the derivative of the Larmor frequency of each transition with respect to $\varepsilon$. The orange colour refers to the out-of-plane $S-T_0$ transition at 12.5 mT. The inset shows a zoom-in at large detunings. (c) Inhomogeneous dephasing time for the $S-T_0$ transition at out-of-plane field as a function of detuning, highlighting that the dephasing time is limited either by Zeeman or detuning noise. (d) Inhomogeneous dephasing time for the $S-T_0$ transition at $\varepsilon= 7.4$ meV as a function of $B_{z}$, demonstrating an insensitivity to the magnetic field value. The error bars correspond to the standard deviation of the fit.}
\label{Fig:Fig4}
\end{figure}

Next, we characterise the inhomogeneous dephasing time $T_2^\ast$ for the measured spin transitions. We begin with an in-plane magnetic field of $30$ mT and measure $T_2^\ast$ as a function of detuning for the three transitions (see Fig.~\ref{Fig:Fig4}(a)). At low detunings, all three transitions exhibit $T_2^\ast$ values around 200 ns. However, with increasing detuning, the $T_--S$  transition (red) shows a coherence increase up to 600 ns, whereas the $T_--T_0$ and $T_--T_+$ transitions extend their dephasing times to the order of several $\mu$s. This variation in trends and values can be attributed to fluctuations in the Larmor frequency relative to $\varepsilon$, i.e. electrical susceptibility $dE/d\varepsilon$. As depicted in Fig.~\ref{Fig:Fig4}(b), the susceptibility for the $T_--T_0$ (purple) and $T_--T_+$ (green) transitions remains close to zero from 5 meV to 8 meV. In contrast, the susceptibility for the $T_--S$ transition (red) only approaches zero at higher detunings. This indicates that the $T_--S$ transition is primarily limited by charge noise, while the $T_--T_0$ and $T_--T_+$ transitions experience two distinct regimes: one dominated by noise in $\varepsilon$ (at low detunings) and another by magnetic fluctuations (at high detunings) ~\cite{wu_two-axis_2014, jirovec_singlet-triplet_2021}. 

Using a simple noise model ~\cite{dial_charge_2013, wu_two-axis_2014}, we fit the three curves in Fig.~\ref{Fig:Fig4}(a) (solid lines) and verify this hypothesis. In this model, $T_2^\ast=\sqrt{2}\hbar/\sqrt{\langle \delta E^2 \rangle}$, where the energy fluctuations $\delta E$ depend on three fitting parameters: $\delta \varepsilon_\text{rms}$, $\delta E_{\Delta Z_\text{rms}}$, and $\delta E_{Z_\text{rms}}$, which represent the root-mean-square (r.m.s.) of the noise on detuning, the Zeeman energy difference, and the total Zeeman energy, respectively (see Supplementary for more details). By considering only the noise contribution from detuning (dashed lines), we confirm that the $T_--S$ dephasing is predominantly caused by detuning noise, while $T_--T_0$ and $T_--T_+$ exhibit two distinct regimes of noise contribution.

We finally perform a similar analysis for the out-of-plane $T_2^\ast$ as a function of $\varepsilon$, as well as for the magnetic field dependence (see Figs.~\ref{Fig:Fig4}(c, d)). As in the in-plane case, dephasing times at low detunings (large exchange) are limited by detuning noise and by magnetic noise at high detunings (see  Fig.~\ref{Fig:Fig4}(c)). In this direction, the qubit quantization axes should be almost co-linear with the hyperfine interaction for heavy-hole qubits, which might explain the larger out-of-plane $\delta E_{\Delta Z_{rms}}$ compared with the in-plane direction (see Table~\ref{table:T2_inplane_params} in Supplementary). Nevertheless, $T_2^\ast$ times reach values of about 400 ns and are unaffected by the magnitude of the magnetic field. \\

\section*{Outlook}

In this work, we exhaustively characterized two hole spins and demonstrated a fully-operational microwave-driven singlet-triplet qubit in a planar germanium heterostructure. Our ac-driving approach enables us to overcome the problem of non-orthogonal control axes, which limits alternative baseband-pulsed singlet-triplet architectures ~\cite{petta_coherent_2005, jirovec_singlet-triplet_2021, zhang_universal_2024, liles_singlet-triplet_2023}. At the same time, low-frequency driving in the MHz regime, renders our qubit less prone to microwave cross-talk, a key issue in the typically GHz-range Loss-DiVincenzo qubit encodings.

A simple protocol, applicable to various geometries and materials —-including those with synthetic SOI due to the presence of micromagnets-— is developed to efficiently characterize all relevant parameters using spectroscopic methods at fixed magnetic field direction. This approach determines the tilt of quantization axes, predicts Rabi frequencies, and identifies different regimes based on the direction of the external magnetic field. Our protocol elucidates the fundamental physics of hole spins while providing insightful information about qubit operation.

In the in-plane regime, we obtain a larger $T_2^\ast$ and faster ac-driving due to a large SOI-induced tilt of quantization axes of the two spins. This large value enables us to probe all possible transitions between different states, and produce the large exchange anisotropy of $\delta\theta=45^\circ$. In addition, we show that $\delta\theta$ is electrically tunable and therefore optimizing $\delta g_\perp$ by controlling the quantum dots shapes and thus g-tensors, e.g. via machine learning algorithms ~\cite{craig_bridging_2024}, will open up the path to leakage-protected two-qubit gates between ST-qubits~\cite{spethmann_high-fidelity_2024, nichol_high-fidelity_2017, zhang_universal_2024}, as well as to more efficient protocols for improving two-qubit gate fidelity between Loss-DiVincenzo qubits~\cite{geyer_anisotropic_2024}. 

In the out-of-plane regime, we also demonstrate a much more uniform quantization axis between the two dots and $\delta\theta\lesssim 5^\circ$, implying an almost isotropic exchange interaction. The properties of this regime enable a larger separation between the computational ($S,T_0$) and non-computational states ($T_\pm$) at low magnetic fields. This separation strongly suppresses leakage, allows uniform readout using conventional Pauli Spin Blockade, and maintains individual qubit addressability. Despite the use of natural germanium, which limits the dephasing time due to the larger effect of hyperfine noise in out-of-plane magnetic fields, we achieve a $T_2^\ast$ time of 400~ns. Importantly, this issue can be straightforwardly resolved in future devices by enrichment of isotopes with zero-spin, a method already common practice for silicon-based qubits. The coherence time is expected to increase one order of magnitude in enriched germanium and become comparable to state-of-the-art hole qubits in silicon~\cite{fischer_spin_2008}. These features along with the ac-driving demonstrated in this work will provide the key elements to scale up ST-based quantum processors~\cite{nichol_high-fidelity_2017, fedele_simultaneous_2021, shulman_demonstration_2012}. 

\section*{Data Availability}
All data included in this work will be available at the Institute of Science and Technology Austria repository.

\section*{Acknowledgements}
We thank A. Crippa for helpful discussions. This research was supported by the Scientific Service Units of ISTA through resources provided by the MIBA Machine Shop and the Nanofabrication facility. This research and related results were made possible with the support of the NOMIS Foundation, the HORIZON-RIA 101069515 project and the FWF Projects with DOI:10.55776/F86 and DOI:10.55776/I5060. M.R.-R. acknowledges support from the Netherlands Organization of Scientific Research (NWO) under Veni
grant VI.Veni.212.223. The Research of S.B. and M.R.-R. was sponsored in part by the Army Research Office and was accomplished under Award Number: W911NF-23-1-0110. The views and conclusions contained in this document are those of the authors and should not be interpreted as representing the official policies, either expressed or implied, of the Army Research Office or the U.S. Government. The U.S. Government is authorized to reproduce and distribute reprints for Government purposes notwithstanding any copyright notation herein.

\section*{Author contribution}

J.S.M fabricated the sample, performed the experiments and analysed the data.  S.C., D.C. and G.I. designed and grew the Ge/SiGe heterostructure.  S. B. developed the theory with input from M. R.-R. .
J. S. M., D. J., Y. S., J. K., M. R.-R., S. B. and G.K. contributed to discussions.  J. S. M., M. R.-R., S. B. and G.K. wrote the manuscript with input from the rest of the authors. G.K. supervised the project.

\section*{Methods}
\subsection{Heterostructure details and device fabrication}

The Ge/SiGe heterostructure used in this work is a $\approx20$ nm-thick strained Ge quantum well capped by a 20 nm layer of Si$_{0.3}$Ge$_{0.7}$. More details can be found in a previous work \cite{jirovec_singlet-triplet_2021}.

The fabrication procedure starts by patterning the ohmic contacts using an electron beam lithography system, followed by argon milling to remove the native SiO$_2$ layer and establish an ohmic contact. Next, 60 nm of Pt is evaporated at a 5$^{\circ}$ angle using an electron beam evaporator. A mesa is then defined by etching away 100 nm in a reactive ion etching process. Subsequently, 15 nm of aluminium oxide is deposited via atomic layer deposition at 300$^{\circ}$C, which anneals the ohmic contacts over approximately 30 minutes. Prior to this, the native SiO$_2$ oxide is removed with a 10-second dip in buffered hydrofluoric acid. Electrostatic gates are patterned on the mesa using electron beam lithography, followed by the deposition of 3 nm of Ti and 20 nm of Pd. The plunger and barrier gates are fabricated in separate steps. Finally, a thick layer of Ti (3 nm) and Pd (127 nm) is used for the bonding pads, which also climb up the mesa to contact the previously defined gates.

\subsection{Experimental setup}
All measurements are performed in a Leiden cryogenics dry dilution cryostat with a base temperature of 50 mK. The fridge is equipped with a two-axis magnet along the Y and Z directions indicated in Fig.~\ref{Fig:Fig1}(a). A detailed sketch of the experimental setup is shown in Fig.~\ref{SFig:Setup}.

\subsection{Theory}

\paragraph{General model.}
We model the double quantum dot close to the transition $(1,1)\to (0,2)$ at detuning $\varepsilon\approx 0$ by a $5\times 5$ extended Fermi-Hubbard Hamiltonian spanning the states $(| S_{02}\rangle,| S_{11}\rangle, |T_{+}\rangle,| T_{-}\rangle $, $|T_{0}\rangle)$ \cite{geyer_anisotropic_2024, ungerer_strong_2024, stepanenko_singlet-triplet_2012} 
\begin{equation}
H = 
\left(
\begin{array}{ccccc}
 \varepsilon & -\sqrt{2}t & 0 & 0 & 0 \\
-\sqrt{2}t & 0 & -\frac{\delta b_x+i\delta b_y}{\sqrt{2}} & \frac{\delta b_x-i\delta b_y}{\sqrt{2}} & \delta b_z \\
 0 & -\frac{\delta b_x-i\delta b_y}{\sqrt{2}} & \bar{b}_z & 0 & \frac{\bar{b}_x-i\bar{b}_y}{\sqrt{2}} \\
  0 & \frac{\delta b_x+i\delta b_y}{\sqrt{2}} & 0 & -\bar{b}_z & \frac{\bar{b}_x+i\bar{b}_y}{\sqrt{2}} \\
 0 & \delta b_z & \frac{\bar{b}_x+i\bar{b}_y}{\sqrt{2}} & \frac{\bar{b}_x-i\bar{b}_y}{\sqrt{2}} & 0 \\
\end{array}
\right) \ ,
\label{eq:H6x6}
\end{equation}
including spin-orbit interactions via spin-flip tunnelling and anisotropies of g tensors.
Following \cite{geyer_anisotropic_2024}, we use the spin-orbit frame, where the spin-flip tunnelling contribution into a local redefinition of the Zeeman energy such that the tunnelling amplitude $t=\sqrt{t_\text{sc}^2+t_\text{sf}^2}$ comprises both spin-conserving $t_\text{sc}$ and spin-flip $t_\text{sf}$ components. We define the difference and total Zeeman energy vectors produced by a magnetic field $\textbf{B}$ as
\begin{align}
\label{eq:deltab}
\delta\textbf{b}&=\frac{\mu_B}{2} \textbf{B} \Big[\underline{g}_L\underline{R}(-\theta_\text{so}/2)-\underline{g}_R\underline{R}(\theta_\text{so}/2) \Big] \ , \\
\label{eq:bbar}
  \bar{\textbf{b}}&=\frac{\mu_B}{2} \textbf{B} \Big[\underline{g}_L\underline{R}(-\theta_\text{so}/2)+\underline{g}_R\underline{R}(\theta_\text{so}/2) \Big]\ ,
\end{align}
where $\underline{g}_{L/R}$ are the symmetric $g$ tensors of the left and right dots, $\underline{R}$ is a rotation matrix around the SOI axis by the spin-flip angle $\theta_\text{so}=2\arctan2(t_\text{sf}, t_\text{sc})\approx 2d/l_\text{so}$, with $d$ being the distance between the dots and $l_\text{so}$ is the spin-orbit length.  In the main text, the definitions of $\underline{g}_1^{\text{so}}$ and $\underline{g}_2^{\text{so}}$ refer to $\underline{g}_L\underline{R}(-\theta_\text{so}/2)$ and $\underline{g}_R\underline{R}(+\theta_\text{so}/2)$, respectively (see Fig.~\ref{SFig:SuppGeo}(a)).

We also consider the microwave driving term coupling to the detuning of the two dots
\begin{align}
\label{eq:coupling}
  H_\text{driving}&= \delta\varepsilon |S_{02}\rangle\langle S_{02}| \ , 
  \end{align}
where $\delta\varepsilon=V_{ac} \frac{d}{dV}\varepsilon $ depends on the susceptibility of the detuning to the potential applied to the gate and on the applied microwave field $V_{ac}$.

We now introduce a more convenient singlet-triplet frame by first diagonalising the singlet sector by including the tunnelling $t$ via a rotation of an angle $\theta=- \text{arctan2}(\varepsilon,\sqrt{2}t_c)/2$. We also fix the direction of the spin quantization axis such that the triplet subsector is diagonal by performing a rotation $\underline{R}_B$ that maps $\bar{\textbf{b}}$ to the $z$-direction, i.e.
\begin{equation}
\underline{R}_B \textbf{n}_z =\bar{\textbf{b}} / \bar{b}  \ , \ \bar{b}=|\bar{\textbf{b}} |
\end{equation}
By introducing the exchange energy $J= (-\varepsilon+\sqrt{\varepsilon^2+8t^2})/ {2}$ and the convenient parametrisation of the rotated vector $\delta{\textbf{b}}$ in terms of  its components perpendicular $\delta b_\perp$ and parallel $\delta b_\parallel$ to $\textbf{b}$ and azimuthal angle $\varphi_B$
\begin{equation}
 \delta{\textbf{b}} \underline{R}_B  =\left( \begin{array}{c}
  \delta b_\perp \cos(\varphi_b)     \\
    \delta b_\perp \sin(\varphi_b)     \\
      \delta b_\parallel            
 \end{array}\right) \ ,
\end{equation}
we find
\begin{equation}
\label{eq:static_H_SM}
\resizebox{\columnwidth}{!}{$
H=
\begin{pmatrix}
J+\varepsilon & 0 & -\frac{\delta b_{\perp}e^{i\varphi_b} }{\sqrt{2}}\cos{\theta} & \frac{\delta b_{\perp}e^{-i\varphi_b}}{\sqrt{2}}\cos{\theta} & \delta b_{\parallel}\cos{\theta}  \\
0 & -J & -\frac{\delta b_{\perp}e^{i\varphi_b}}{\sqrt{2}}\sin{\theta} &  \frac{\delta b_{\perp}e^{-i\varphi_b}}{\sqrt{2}}\sin{\theta} & \delta b_{\parallel} \sin{\theta}  \\
-\frac{\delta b_{\perp}e^{-i\varphi_b}}{\sqrt{2}}\cos{\theta} & -\frac{\delta b_{\perp}e^{-i\varphi_b}}{\sqrt{2}}\sin{\theta} & \bar{b} & 0 & 0 \\
 \frac{\delta b_{\perp}e^{i\varphi_b}}{\sqrt{2}}\cos{\theta} & \frac{\delta b_{\perp}e^{i\varphi_b}}{\sqrt{2}}\sin{\theta} & 0 & -\bar{b}& 0 \\
\delta b_{\parallel} \cos{\theta} & \delta b_{\parallel} \sin{\theta}  & 0 & 0 & 0 \\
\end{pmatrix}
$} \ ,
\end{equation}
and 
\begin{equation}
\label{eq:H-driving}
H_\text{driving}= -\delta\varepsilon\left(
\begin{array}{ccccc}
 \sin ^2(\theta ) & -\sin (2\theta )/2 & 0 & 0 & 0 \\
 -\sin (2\theta )/2 & \cos ^2(\theta ) & 0 & 0 & 0 \\
 0 & 0 & 0 & 0 & 0 \\
 0 & 0 & 0 & 0 & 0 \\
 0 & 0 & 0 & 0 & 0 \\
\end{array}
\right) \ .
\end{equation}

These equations correspond to Eq.~\eqref{eq:static_H} and $H_\text{driving}$ in the main text, where the state $| S_{02}\rangle$ was not included as its energy contribution can be neglected. As the azimuthal angle $\varphi_B$ has no impact on the spectrum and amplitudes to the lowest order, there we set it to $\varphi_B=0$.

For convenience of notation, we introduce also the dimensionless g-factors associated to these Zeeman vector components as
\begin{align}
\bar{g} &=\bar{b}/\mu_B B \ , \\
\delta g_{\perp,\parallel}&=\delta b_{\perp,\parallel}/\mu_B B
\end{align}

We use the model in Eq.~\eqref{eq:static_H_SM} to extract the main spectroscopic features in this experiment.  Full numerical diagonalisation is straightforward and produces the energy diagram in Fig.~\ref{Fig:Fig1}(e). In addition, we simulate amplitudes of the transitions between the different states shown in the main text by projecting the driving term in Eq.~\eqref{eq:H-driving}  onto the eigenstates of  Eq.~\eqref{eq:static_H_SM}.

To interpret these results more intuitively, we consider here a few limiting cases.
We note that in the (1,1) sector, one can neglect the effect of the singlet at high energy $J+\varepsilon$ in Eq.~\eqref{eq:static_H_SM}, and we neglect this state in the simplified theory discussed below. 
\paragraph{Large $J$.}
At large $J$ and low B field, there is an anticrossing of $S-T_-$ states at $J=\bar{b}$, with amplitude $\Delta_{ST}=\sqrt{2}\delta b_\perp\sin(\theta)\approx \sqrt{2}\delta b_\perp$. The approximation $\sin(\theta)\approx 1$ is valid deep in the (1,1) sector (positive values of $\varepsilon$) for $t\ll |\varepsilon|$.

By accounting for the hybridisation of $S-T_-$ states, we find  that to first order in the remaining $\delta \textbf{b}$'s the transition energies from the ground state to the excited states are 
\begin{align}
\varepsilon_1&=\sqrt{(J-\bar{b})^2+2\delta b_\perp^2} \ , \\
\varepsilon_2&= \frac{J+\bar{b}+\sqrt{(J-\bar{b})^2+2\delta b_\perp^2}}{2}\ , \\
\varepsilon_3&=\bar{b}+\frac{J+\bar{b}+\sqrt{(J-\bar{b})^2+2\delta b_\perp^2}}{2} \ .
\end{align}
We note that in this regime, $\delta b_\parallel$ only corrects these transitions energies to the second order with respect to $J~\bar{b}$.

Including the driving Hamiltonian $H_\text{driving}$ in Eq.\eqref{eq:H-driving}, and by standard perturbation theory in $\delta \textbf{b}$, we also find  the amplitudes from the transition matrix elements $A_i=\bra{i}H_\text{driving}\ket{g}$ of the ground state to the $i$-th state in the (1,1) sector as
\begin{align}
\label{eqs:Rabis1}
A_1 &\approx \frac{\delta\varepsilon}{|\varepsilon|}  \frac{\delta b_\perp}{\sqrt{2}}   \frac{J}{\sqrt{(J-\bar{b})^2+2\delta b_\perp^2}} \ , \\
\label{eqs:Rabis2}
A_2 &\approx \frac{\delta\varepsilon}{|\varepsilon|} \frac{\delta b_\parallel}{\sqrt{2}}   \sqrt{1+\frac{J-\bar{b}}{\sqrt{(J-\bar{b})^2+2\delta b_\perp^2}} }\ , \\
\label{eqs:Rabis3}
{A_3} & \approx \frac{\delta\varepsilon}{|\varepsilon|} \frac{ \delta b_\perp}{{2} (1+\bar{b}/J)}   \sqrt{1+\frac{J-\bar{b}}{\sqrt{(J-\bar{b})^2+2\delta b_\perp^2}} }  \ .
\end{align}

We note the second order coupling terms $\delta\varepsilon \delta b_\perp$ and $\delta\varepsilon \delta b_\parallel$ drive the Rabi oscillations despite the diagonal structure of the triplet sector in the Hamiltonian.

\paragraph{Small $J$.}
When the magnetic field is large, it is convenient to change frame. Following ~\cite{geyer_anisotropic_2024}, we move to the qubit frame, defined by diagonalizing the individual qubits. In this frame, we find the two-qubit Hamiltonian with anisotropic exchange interaction
\begin{align}
&H_{Q}= \frac{b_1}{2}  \sigma_z^{(1)}+\frac{b_2}{2} \sigma_z^{(2)}+\frac{J}{4} \left[\pmb{\sigma}^{(1)} \underline{R}_y(-\delta\theta) \pmb{\sigma}^{(2)}- \cos(\delta\theta)\right] \\
&=\left(
\begin{array}{cccc}
 \frac{b_1+b_2}{2} & \frac{J\sin\delta \theta }{4} & \frac{-J\sin\delta \theta}{4} & \frac{-J \sin ^2\frac{\delta \theta }{2}}{2}  \\
 \frac{J \sin \delta \theta }{4}  & \frac{b_1-b_2-J \cos \delta \theta }{2} & \frac{J \cos ^2\frac{\delta \theta }{2}}{2}  & \frac{J \sin \delta \theta }{4} \\
 \frac{-J \sin \delta \theta}{4} & \frac{J \cos ^2\frac{\delta \theta }{2}}{2}  & \frac{b_2-b_1-J \cos \delta \theta }{2}  & \frac{-J \sin \delta \theta }{4} \\
 \frac{-J \sin ^2\frac{\delta \theta }{2}}{2} & \frac{J \sin \delta \theta }{4}  & \frac{-J \sin \delta \theta}{4}  & \frac{-b_1-b_2}{2} \\
\end{array}
\right) ,
\end{align}
where the qubits energies are
\begin{align}
b_1&\approx\mu_B B \sqrt{(\bar{g}+\delta g_\parallel )^2+\delta g_\perp^2} \ ,\\
 b_2&\approx \mu_B B  \sqrt{(\bar{g}-\delta g_\parallel)^2+\delta g_\perp^2} \ .
\end{align}
The exchange matrix 
\begin{equation}
J \underline{R}_y(-\delta\theta)=  J  \left(
\begin{array}{ccc}
 \cos (\delta \theta ) & 0 & -\sin (\delta \theta ) \\
 0 & 1 & 0 \\
 \sin (\delta \theta ) & 0 & \cos (\delta \theta ) \\
\end{array}
\right) \ ,
\end{equation}
 comprises a rotation with respect to the $y$-direction of the angle $\delta\theta$, that is directly extracted from the parameters measured in our work as 
\begin{align}
\label{eq:rotationJmatrix}
\delta\theta &= \text{arctan}\left[\frac{\delta b_\perp\sin\theta}{\bar{b}+ \delta b_\parallel\sin\theta}\right]+\text{arctan}\left[\frac{\delta b_\perp\sin\theta}{\bar{b}- \delta b_\parallel\sin\theta}\right] \\
&\approx \text{arctan}\left[\frac{\delta g_\perp}{\bar{g}+  \delta g_\parallel}\right]+\text{arctan}\left[\frac{\delta g_\perp}{\bar{g}- \delta g_\parallel}\right] \ . 
\end{align}

The approximate signs here indicate that we neglected the  $\sin(\theta)\approx 1$ correction that weights the $\delta \textbf{b}$'s, and introduce a weak detuning dependence in these parameters.

From the parameters extracted in this work, we find that the anisotropic angle of the exchange is 45 degree in-plane and 5 degrees in the out-of-plane direction. 
We note that the isotropic exchange case reached at $\delta\theta=0$ is the native interaction for SWAP operations, while the completely anisotropic case $\delta\theta=90^\circ$ is optimal for resonant CNOT gates between two Loss-DiVincenzo gates ~\cite{geyer_anisotropic_2024} as well as for suppressing leakage in CZ gates between ST qubits ~\cite{spethmann_high-fidelity_2024}.

In this limit, the triplet $T_-$ is the ground state and when the $\delta \textbf{b}$'s are sufficiently large the transitions energies  are 
\begin{align}
\varepsilon_1 &=b_2 -\frac{J}{2}\cos\delta\theta \ , \\
\varepsilon_2&=b_1 -\frac{J}{2}\cos\delta\theta \ , \\
\varepsilon_3 &=b_1+b_2 \ .
\end{align}
Moreover, focusing on the transitions generated by the $H_\text{driving}$ from the $T_-$ ground state to the excited states and keeping only the lowest order terms, we find the amplitudes
\begin{align}
A_1&\approx \frac{\delta\varepsilon}{|\varepsilon|}\sin\delta\theta \left(\frac{1}{J}+\frac{b_1^3 \cos \delta \theta +2 b_1^2 b_2-b_2^3}{2 b_1 b_2(b_1^2 - b_2^2)} \right) \ , \\
A_2&\approx \frac{\delta\varepsilon}{|\varepsilon|}\sin\delta\theta \left(\frac{1}{J}-\frac{b_2^3 \cos \delta \theta +2 b_1 b_2^2-b_1^3}{2 b_1 b_2(b_1^2 - b_2^2)} \right) \ , \\
A_3 &\approx \frac{\delta\varepsilon}{|\varepsilon|} \frac{2}{J} \sin^2\left(\frac{\delta\theta}{2} \right) \ .
\end{align}

\paragraph{Protocol to extract relevant parameters.}

We now discuss a protocol to identify the relevant experimental parameters from the magnetic field sweep in Fig.~\ref{Fig:Fig1}(c) and~\ref{Fig:Fig3}(a). Here, we consider a fixed value of detuning $\varepsilon$ and of magnetic field direction, and vary only the amplitude of $B$.

We first extract $J$  by considering that at $B=0$, the ground state is the low energy singlet and is split from the three degenerate triplets by $J$. 

We then extract $\bar{g}$ and $\delta g_\perp$ by focusing on the $S-T_-$ anticrossing. For small values of $\delta g_{\perp,\parallel}$, the anticrossing occurs at the magnetic field $B^*$ where $\bar{b}=J$, from which we find 

\begin{equation}
    \bar{g}=J/\mu_B B^* \ .
\end{equation} 

To the lowest order $\delta \textbf{b}$, the amplitude of the anticrossing is 
\begin{equation}
    \Delta_{ST}\approx\sqrt{2}\delta b_\perp \to \delta g_\perp= \Delta_{ST}/\sqrt{2} \mu_B B^* \ ,
\end{equation}
where we neglected the small correction $\sin(\theta)\approx 1$ deep in the (1,1) sector where $t\ll |\varepsilon|$ .

Finally, we extract $\delta g_\parallel$  from spectroscopic features at large $B$ fields. For example in particular, the slope $\partial_B \varepsilon_{1,2}$ of the first two low energy transitions from $T_-$ to the first and second excited states with respect to $B$ is related to $\delta g_\parallel$ as
\begin{equation}
   \delta g_\parallel=\left|\sqrt{(\partial_B \varepsilon_{1,2}/\mu_B)^2-\delta g_\perp^2} - \bar{g} \right|\ .
\end{equation}

\bibliographystyle{naturemag}
\bibliography{Bibliography}

\clearpage
\onecolumngrid 

\section*{Supplementary information}
\setcounter{figure}{0}

\renewcommand\thefigure{S\arabic{figure}}
\renewcommand{\theHfigure}{\thefigure}

\subsection*{Experimental setup.}

\begin{figure}[H]
    \centering
    \includegraphics[scale=0.7]{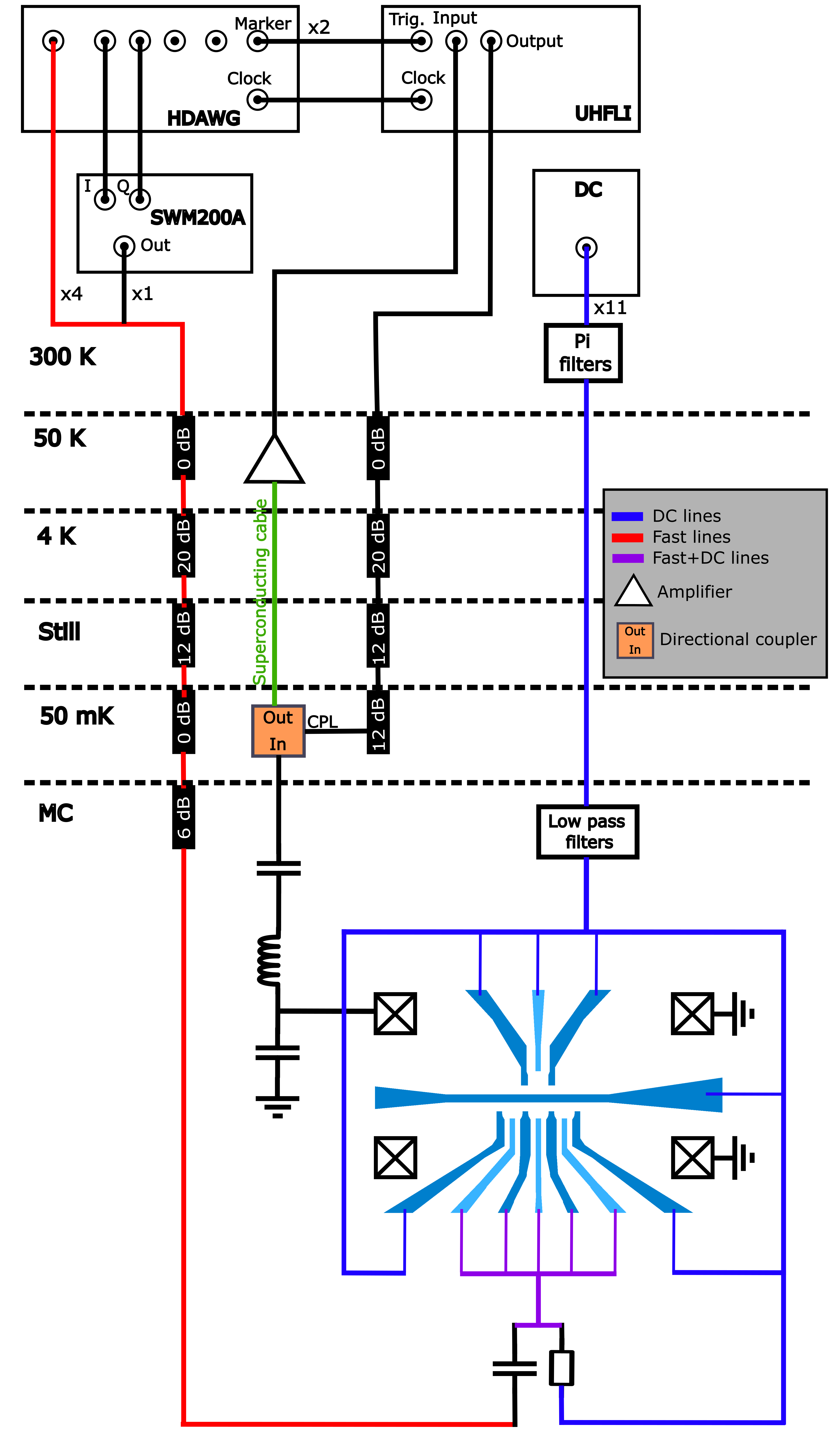}
    \caption{Sketch of experimental setup. High-frequency lines (red color) are attenuated at the different stages of the cryostat to minimize electron temperature. The attenuators  are represented as black boxes with the correspondent values. A 0 dB attenuator acts as a thermal anchor. The fast lines are connected to an Arbitrary Waveform Generator from Zurich Instruments (HDAWG) to change the charge occupation of the qubit, and one line is connected to a Rhode Schwarz microwave source (SMW200A) to drive the qubit. The DC lines (in blue) are connected to a Delft-IVVI rack. 
    The UHFLI generates and demodulates the microwave reflectometry signal. This reflectometry tone is attenuated and applied to the ohmic contact of the charge sensor with a directional coupler (MiniCircuits ZFDC-20-50 S+). The tank circuit used for reflectometry comprises an inductance (L) of 470 nH and an approximate parasitic capacitance ($C_p$) of 0.9 pF. The reflected signal is amplified at the 50 K stage by a cryogenic amplifier (CITLF3).}
    \label{SFig:Setup}
\end{figure}

\subsection*{Geometrical interpretation.}

\begin{figure}[H]
    \centering
    \includegraphics[scale=1.1]{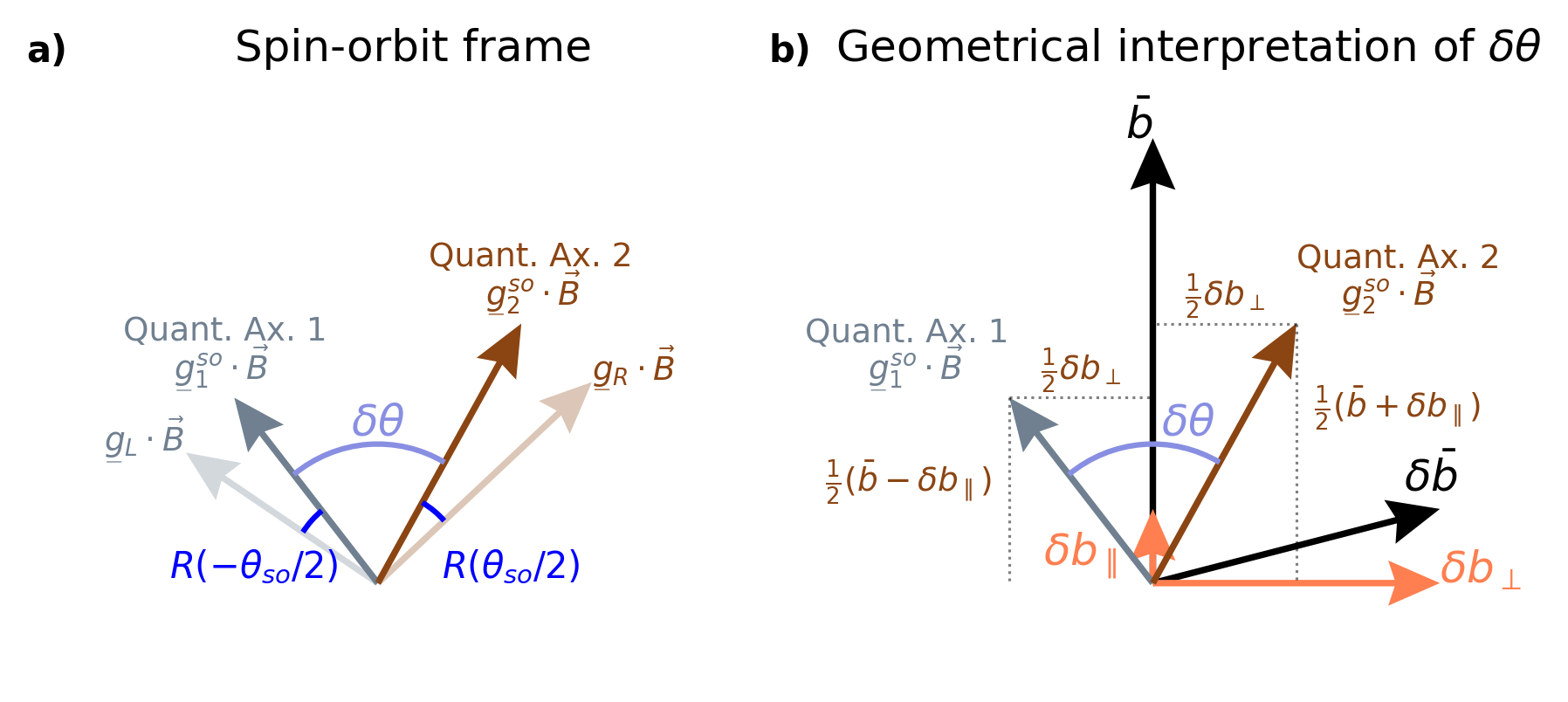}
    \caption{(a) Schematic explaining the difference between the laboratory and the spin-orbit frames. The light grey/brown arrows correspond to the g-tensors of left/right quantum dots, respectively. The dark grey/brown arrows represent the quantization axes in the spin-orbit frame, where the light quantization axes have been rotated by an angle $\theta_{\text{so}}/2$ in order to take into account the spin-flip tunneling. The angle $\delta \theta$ between quantization axes 1 and 2 considers the spin-flip tunneling and the tilt between the g-tensors $\underline{g}_L$ and $\underline{g}_R$. (b) Geometrical explanation of $\delta\theta$ using the parameters extracted from the spectroscopy as a function of the magnetic field. By applying the trigonometric relations one can prove that $\delta\theta=\text{arctan}\left[\delta g_\perp/(\bar{g}+ \delta g_\parallel)\right]+\text{arctan}\left[\delta g_\perp/(\bar{g}- \delta g_\parallel)\right]$.}
    \label{SFig:SuppGeo}
\end{figure}

\clearpage

\subsection*{Rabi chevrons and gate fidelities.}
Here we show Rabi chevrons of the three studied transitions in the in-plane magnetic field direction and the estimation of the gate fidelities.
\begin{figure}[H]
\centering
\includegraphics[width=\linewidth]{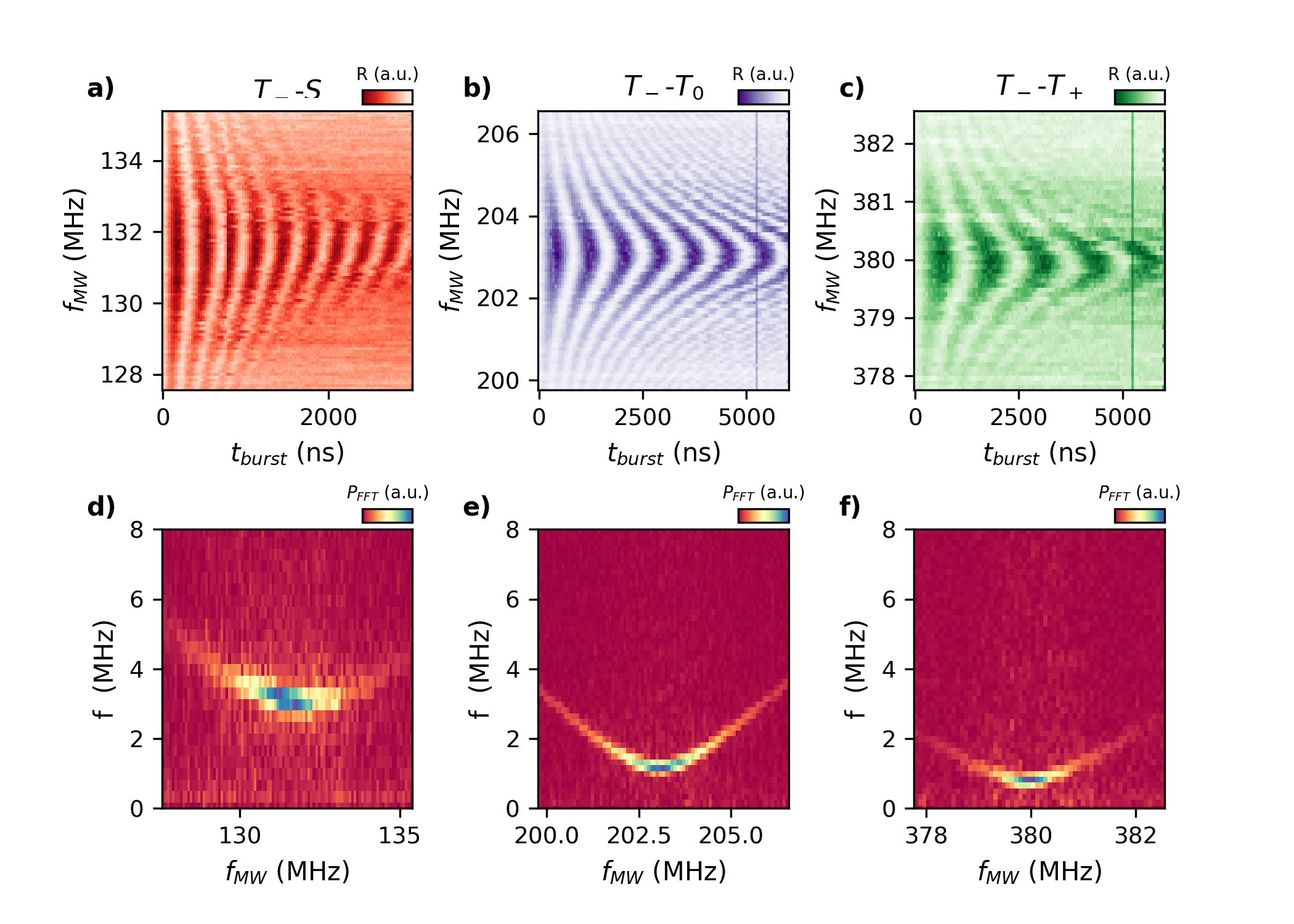}
\caption{(a-c) Rabi chevron for the 3 transitions at $B_y=40$ mT and $\varepsilon=4.0$ meV. (d-f) Fast Fourier transformation for the Rabi Chevrons. The distinction between the $T_-$ and $T_+$ states is made possible by the Landau-Zener transition probability when returning to the readout point. In our system, the $T_-$ state in the effective (1,1) configuration maps into an unblocked state due to the anticrossing, while the other states remain blocked. 
}
\label{SFig:Supp_RabiChevrons}
\end{figure}

\begin{figure}[H]
    \centering   \includegraphics[scale=0.6]{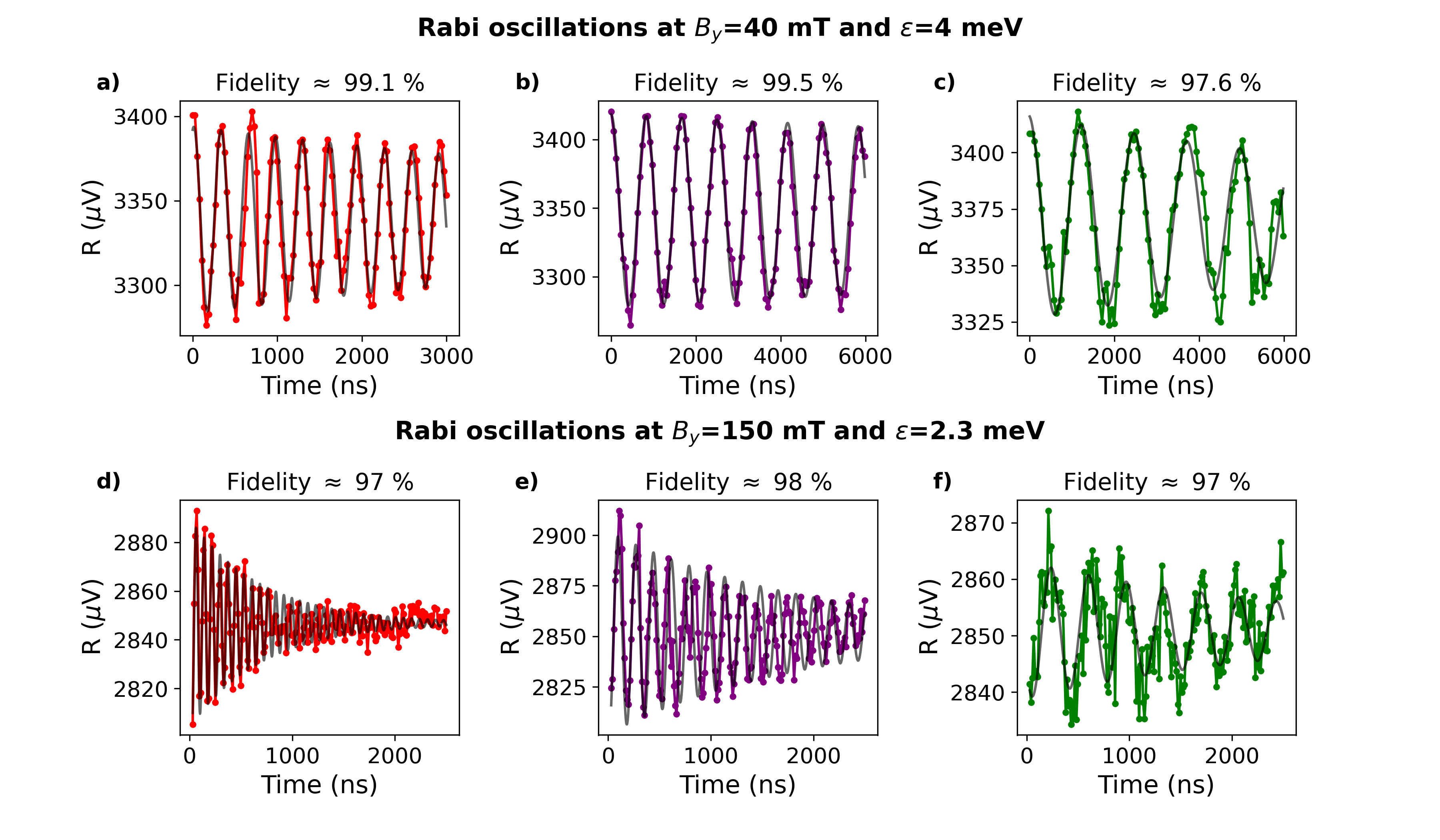}
    \caption{Rabi oscillations for the three transitions investigated at different magnetic field and detuning point. Each trace is fitted to: $A \cos(2\pi f_R t + \phi) \exp(-t/T_2^{\text{Rabi}}) + B$. Data (a-c) are the resonant traces from the Rabi chevrons in Fig.\ref{SFig:Supp_RabiChevrons}. (d-f) are the line traces at 150 mT from the data shown in Fig. \ref{SFig:SupRabis25mV}.The gate fidelity of each transition is estimated using $F =1/2(1+exp(- T_2^{\text{Rabi}}f_R/2))$, where $T_2^{\text{Rabi}}$ the Rabi decay and $f_R$ the Rabi frequency~\cite{stano_review_2022}.}
    \label{fig:EstimatedFidelity}
\end{figure}

\newpage
\subsection*{Raw data of Rabi oscillation dependences.}
In this section, we present the raw data of the measured Rabi oscillations as a function of magnetic field and detuning. For each measurement, the corresponding fast Fourier transform is shown below, revealing the dependence of the Rabi frequency. We also show an extra dataset of Rabi frequencies for the three transitions as a function of detuning and magnetic field in Fig.~\ref{SFig:SupSims45mVand20mT}. These plots were measured using a different magnetic field and detuning configuration compared to Fig.~\ref{Fig:Fig2} in the main text, demonstrating that the model works for different parameters.

\begin{figure}[H]
\centering
\includegraphics[width=\linewidth]{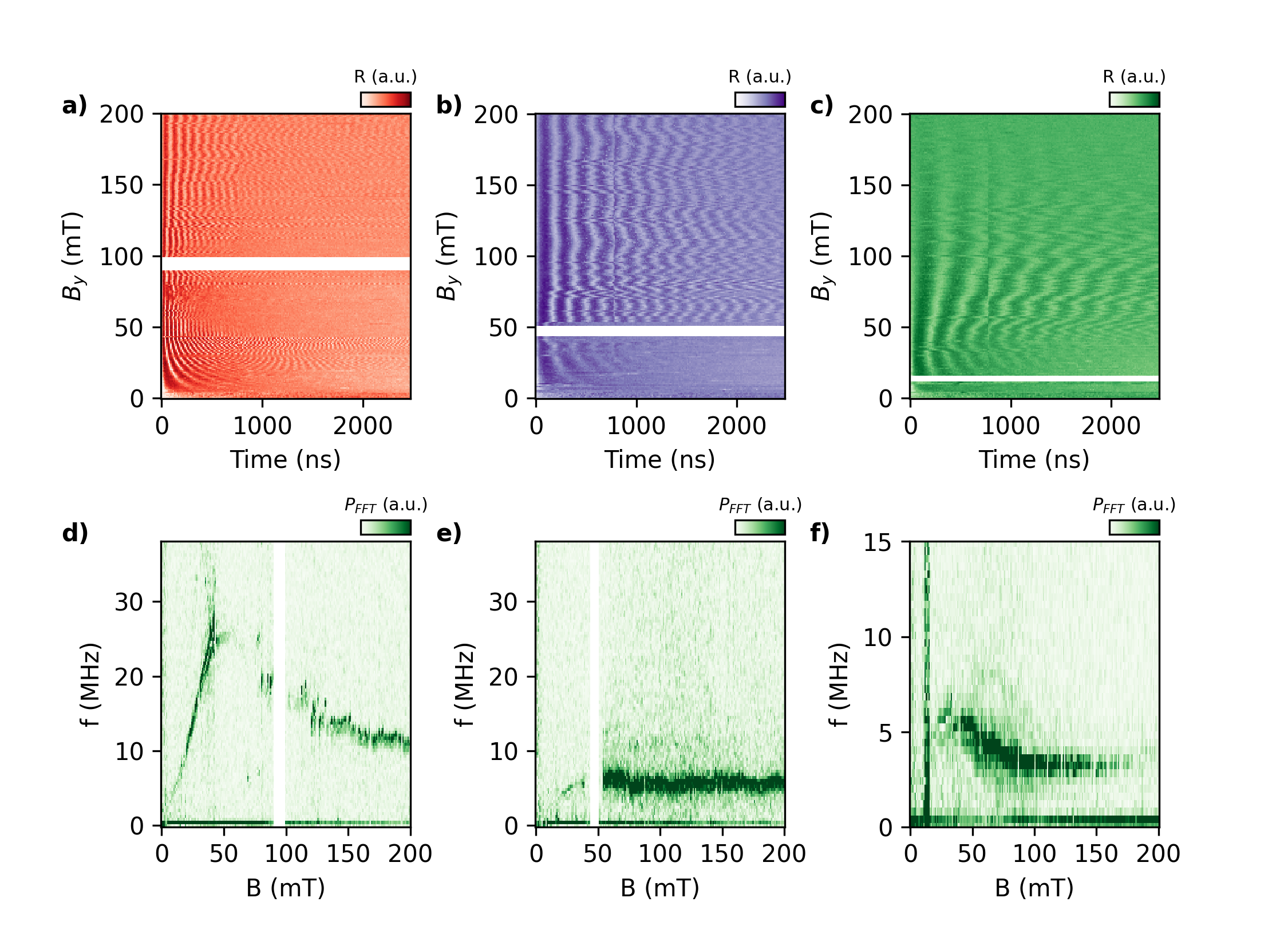}
\caption{Panels (a-c) show Rabi oscillations at $\varepsilon=2.34$ meV as a function of magnetic field $B_y$ for the three transitions. Panels (d-f) show the fast Fourier transform (FFT) of the top panels. The magnetic field-independent signal observed at low frequencies (below 1 MHz) is an artifact of the FFT, resulting from the decay of the Rabi oscillations.}
\label{SFig:SupRabis25mV}
\end{figure}

\begin{figure}[H]
\centering
\includegraphics[width=\linewidth]{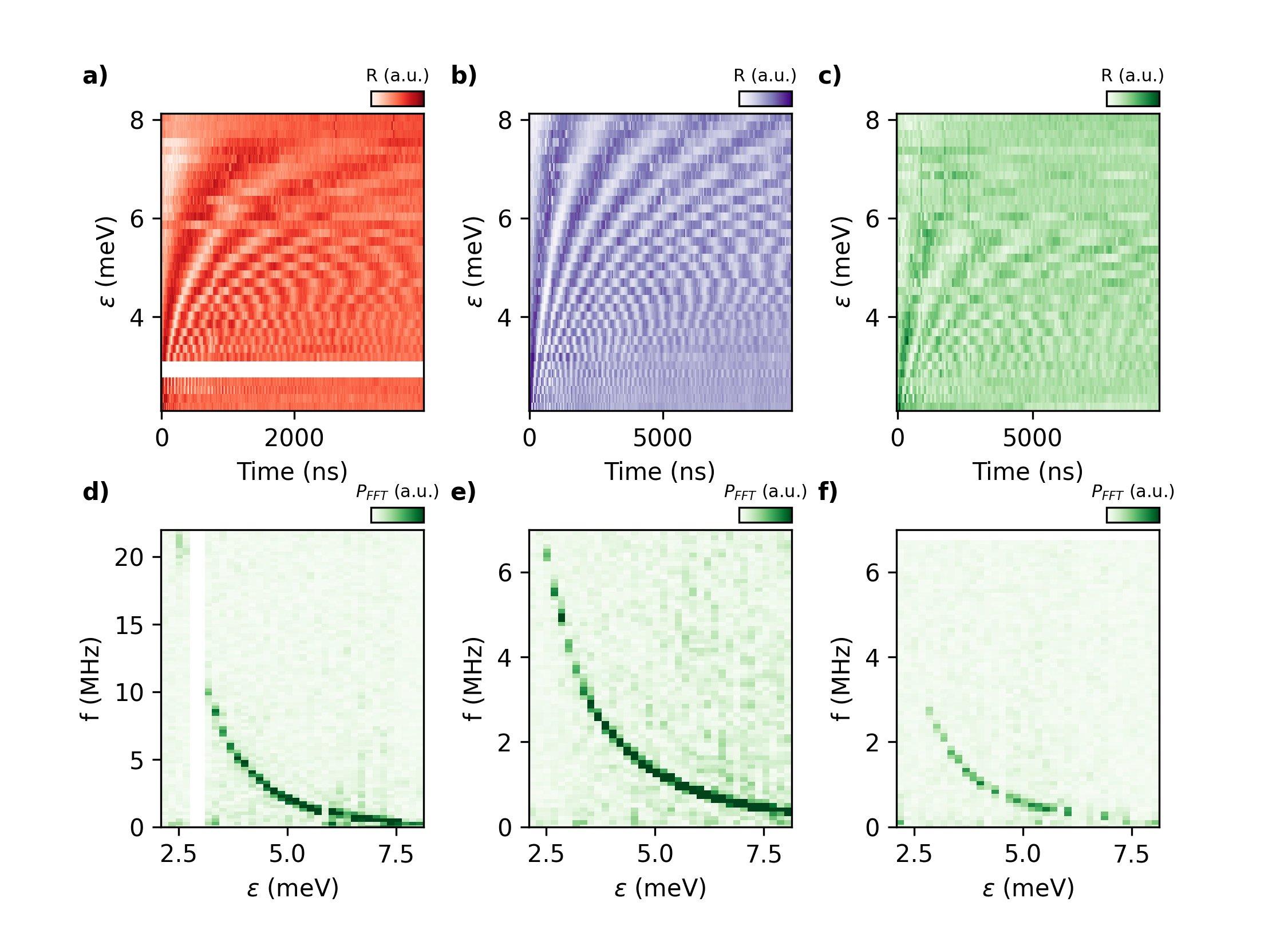}
\caption{Panels (a-c) show Rabi oscillations at $B_y=90$ mT as a function of $\varepsilon$ for the three transitions. Panels (d-f) show the fast Fourier transform (FFT) of the top panels. }
\label{SFig:SupRabis90mT}
\end{figure}

\begin{figure}[H]
\centering
\includegraphics[width=\linewidth]{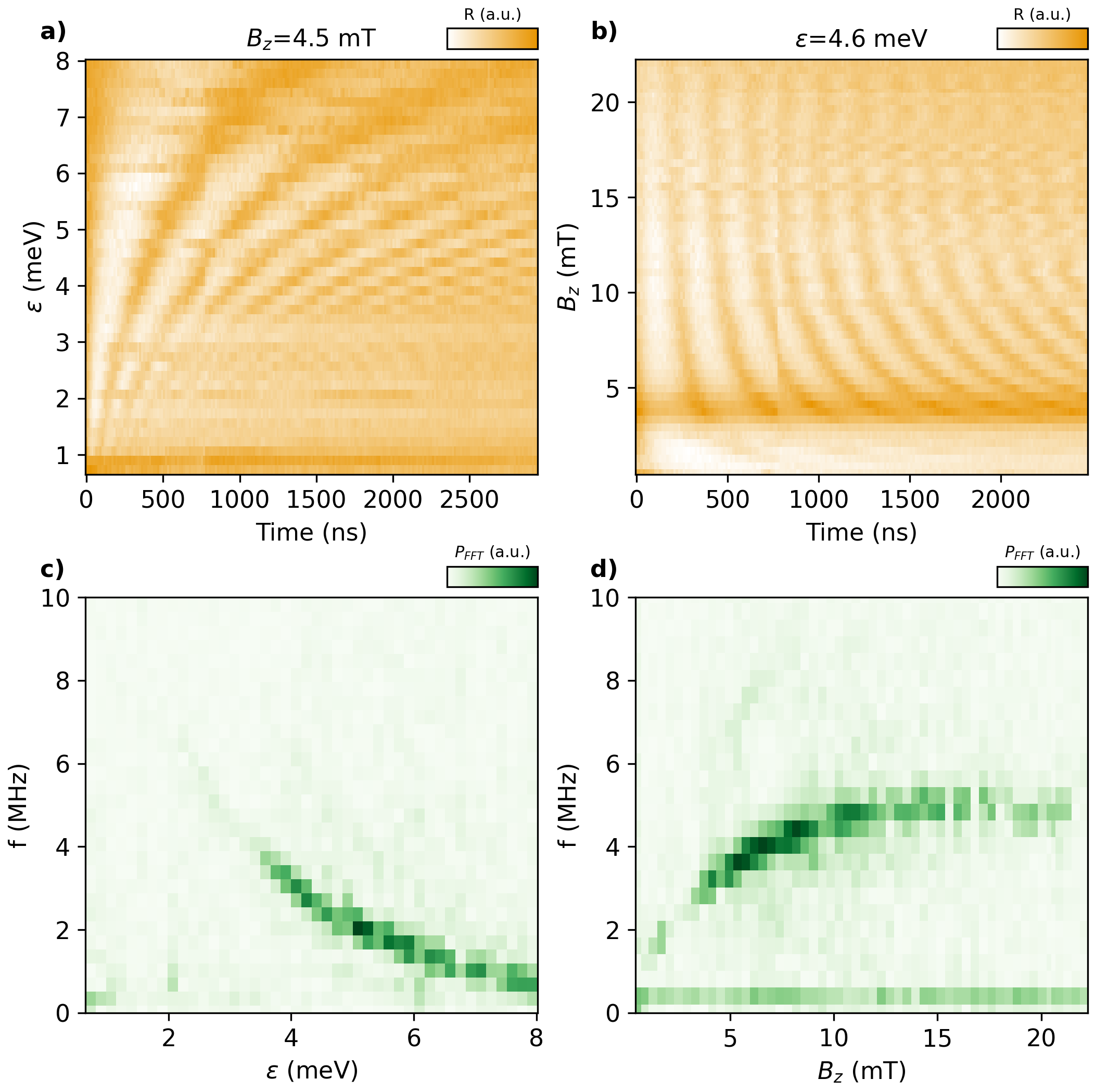}
\caption{Panels a) and b) show Rabi oscillations out-of-plane as a function of detuning and magnetic field, respectively. Panels c) and d) show the respective fast Fourier transforms of the top panels. This is the data used for Fig.~\ref{Fig:Fig3}(c) and (d).}
\label{SFig:SuppST0}
\end{figure}

\begin{figure}[H]
\centering
\includegraphics[width=\linewidth]{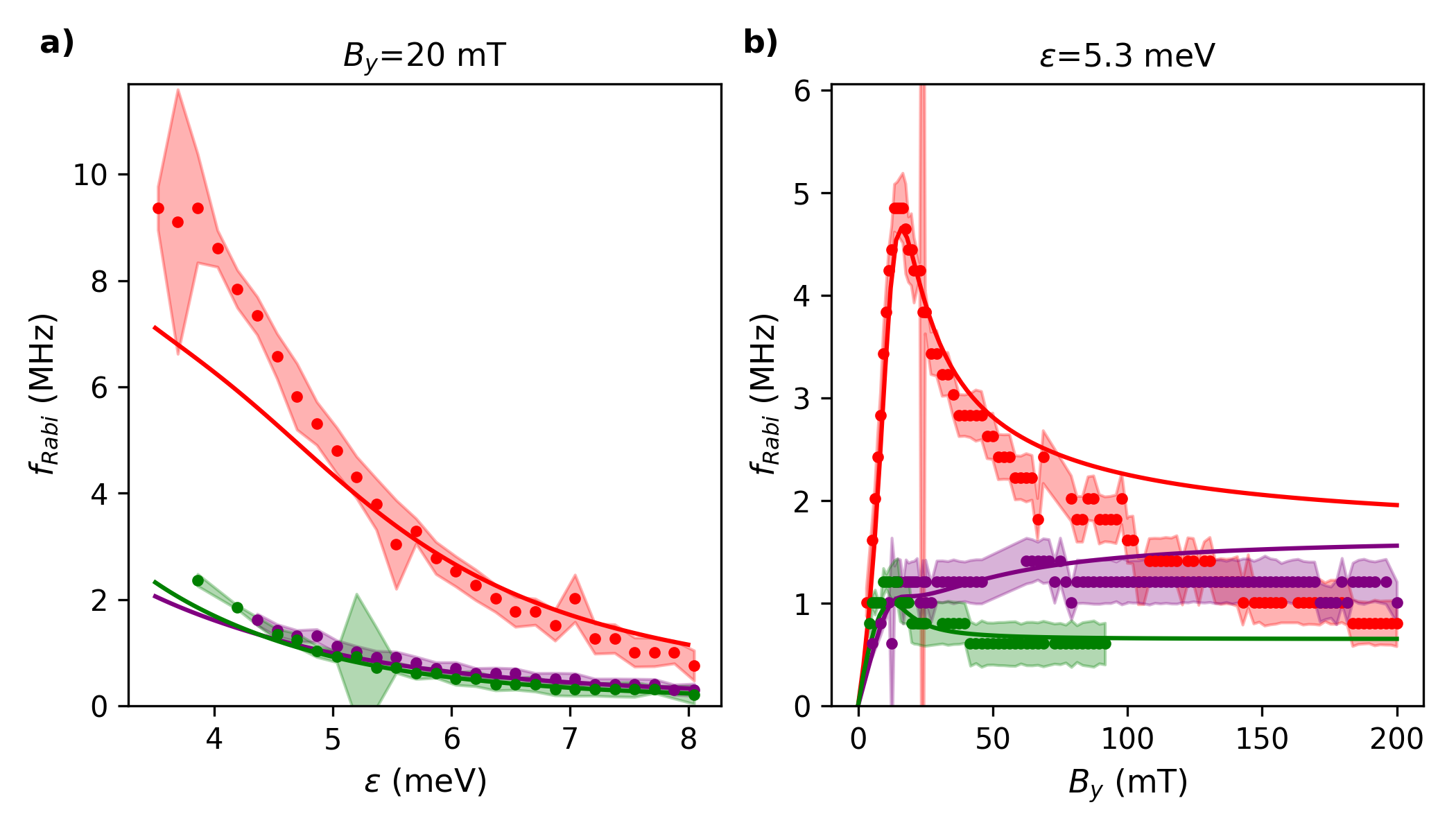}
\caption{Rabi frequencies of the three transitions as a function of detuning (a) and magnetic field (b). The exact solution for the Rabi frequencies is shown with solid lines. These plots have been measured for a different magnetic field and detuning configuration compared with Fig.~\ref{Fig:Fig2} of the main text. }
\label{SFig:SupSims45mVand20mT}
\end{figure}

\subsection*{Lever arms and cross-capacitance of gate BR.}
\begin{figure}[H]
\centering
\includegraphics[width=\linewidth]{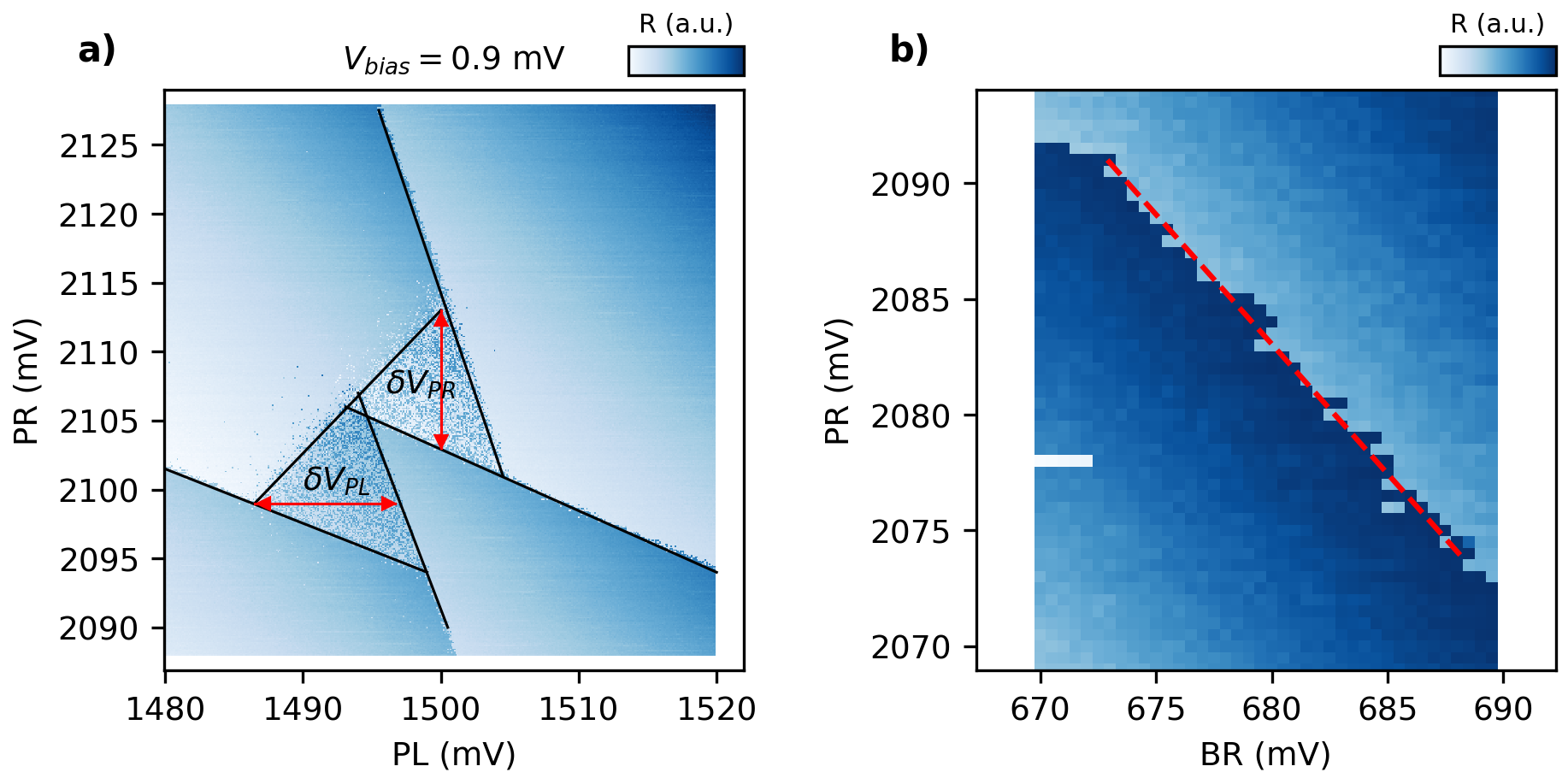}
\caption{a) Bias triangles at a bias voltage of 0.9 mV measured in reflectometry. From $\delta V_{PL}=11.7$ mV and $\delta V_{PR}=10.8$ mV we extract the left (right) lever arm $\alpha_{PL}=0.083$ eV/V ($\alpha_{PR}=0.077$ eV/V). b) Shift of the electrochemical potential in the right quantum dot due to the cross-capacitance from the barrier right (BR). From the slope, indicated with the dashed line, we extract a cross-capacitance $\gamma_{BR-PR}= 1.1$, which implies that BR has a stronger effect on the electrochemical potential of the right QD than the plunger gate. From $\alpha_{PR}$ and $\gamma_{BR-PR}$ we estimate the effect of the microwave burst applied on BR has on the detuning.}
\label{SFig:SuppLeverArms}
\end{figure}

\subsection*{Extra out-of-plane data.}
\begin{figure}[H]
\centering
\includegraphics[width=\linewidth]{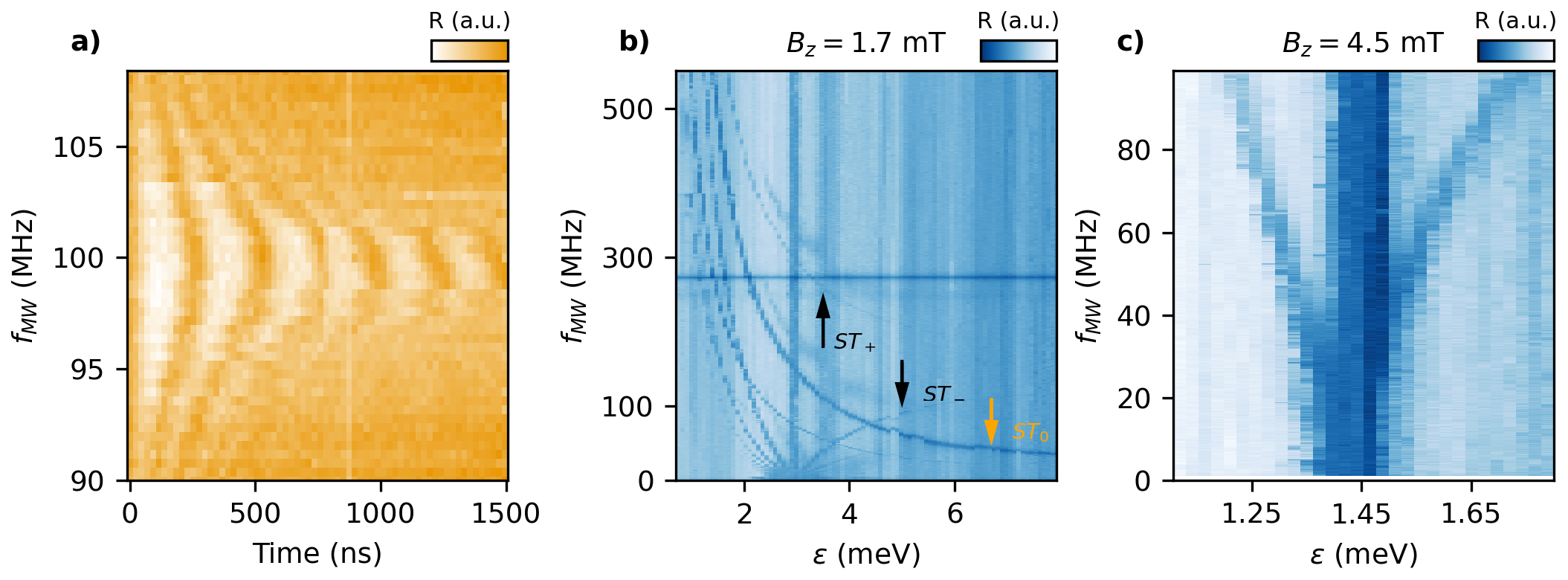}
\caption{Additional measurements in the out-of-plane direction at 7 mT and $\varepsilon=4.0$ meV. a) Rabi Chevron showing microwave driven $S-T_0$ oscillations. b) Spectroscopy revealing the evolution of the three transistions with detuning. c) Zoom-in at low powers of the crossing between the S and $T_-$ states. The change in the background signal at 1.45 meV does not allow resolving the merging point of the two lines and therefore whether a crossing or an anticrossing takes place, but it allows to set the upper bound for the anticrossing size at 30 MHz. The fact that there is a blocked return probability at 1.45 meV demonstrates that there is an anticrossing where the $S$ and $T_-$ states mix.} 
\label{SFig:SupOutPlane}
\end{figure}

\subsection*{Subharmonic processes.}
In this section, we provide additional data shining light into the origin of the integer fractions of the Larmor frequency, also referred as the subharmonic transitions. Such transitions have been also characterized in Refs.~\cite{stehlik_extreme_2014, scarlino_second-harmonic_2015}. Panel (a) presents the spectrum at $B_y = 70$ mT as a function of the microwave power. At low power levels, only the $T_- - S$ and $T_- - T_0$ transitions are observed (indicated by solid red and purple lines, respectively). As the power increases, the $T_- - T_+$ transition emerges, followed by subharmonics of the other two transitions (dashed lines). Panel (b) shows the same dependence but at a lower detuning, corresponding to a larger exchange interaction. Here, due to the highly nonlinear behavior of $J$ at such low detuning, subharmonic effects appear at lower powers, allowing the observation of more transitions. 
We next focus on studying the $T_- - T_0$ transition for the fundamental frequency and its first subharmonic, also referred to in the literature as "one-photon" and "two-photon" processes. Panels (c) and (d) demonstrate that the $T_- - T_0$ transition can be coherently driven by both one-photon and two-photon processes. By analyzing the power dependence of these transitions, we find that the Rabi frequencies follow a power-law relationship, where the exponent corresponds to the number of photons involved in the process, as previously reported in~\cite{scarlino_second-harmonic_2015}.

\newpage

\begin{figure}[H]
\centering
\includegraphics[width=\linewidth]{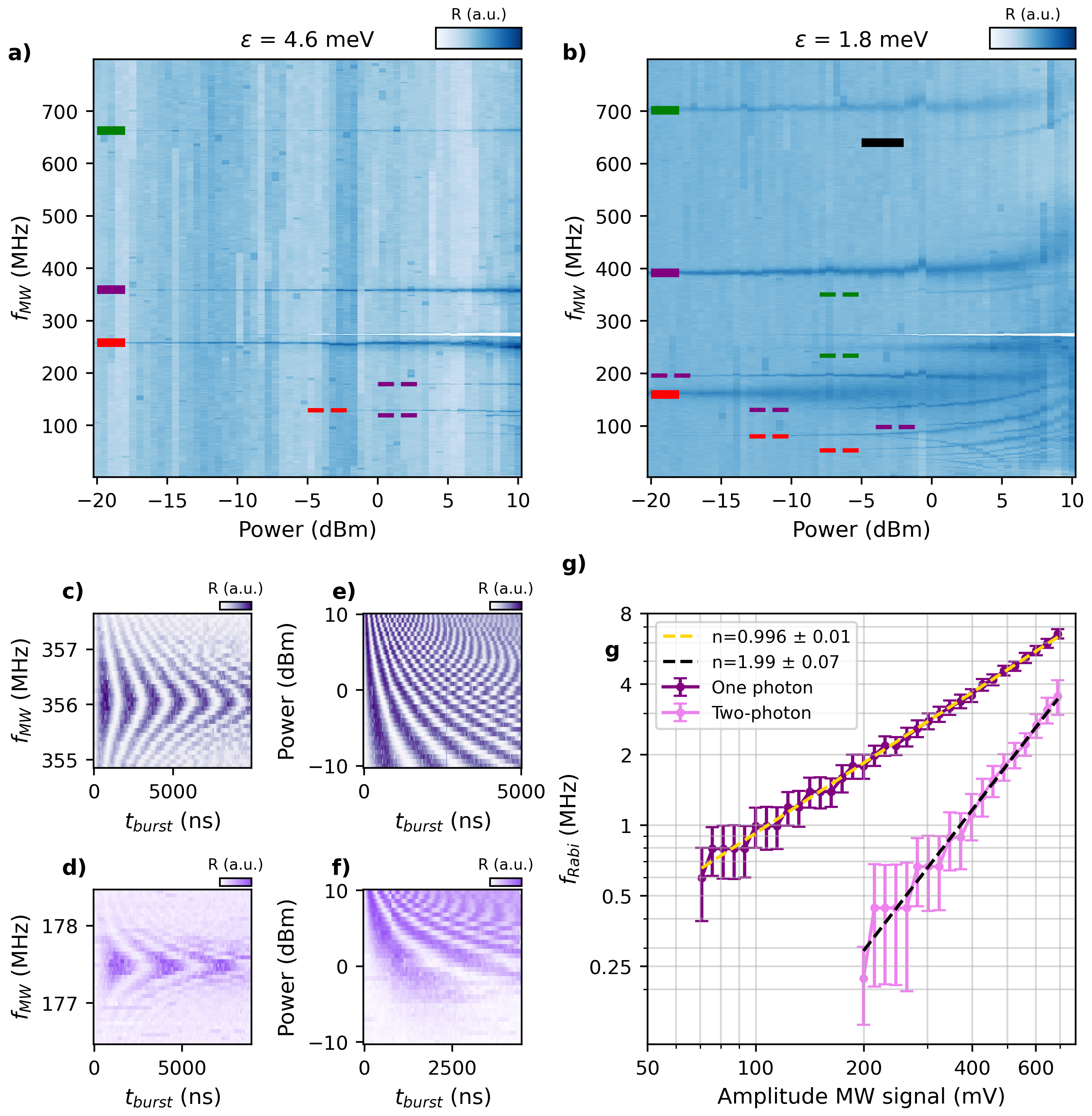}
\caption{Spin transitions as a function of the burst power at $B_y =70$ mT for a detuning of 4.6 meV in (a) and 1.8 meV (b). Solid coloured lines indicate the larmor frequency of the transitions and dashed coloured lines the subharmonic transitions. The black solid line indicates a transition which only appears at low detunings and high powers. (c) Rabi Chevron of transition $T_--  T_0$ driven at the Larmor frequency. (d) Rabi Chevron of transition $T_--T_0$ driven at the half value Larmor frequency ("two-photon process"). Power dependence of Rabi oscillations driven by (e) "one-photon" process and (f) "two-photon" process. (g) Fast Fourier Transform of (e) and (f) showing the Rabi frequency dependence as a function of the amplitude of the microwave signal.  }
\label{SFig:SuppHarmonics}
\end{figure}

\newpage
\subsection*{Noise model and fitting parameters}

Following the noise model introduced in reference \cite{dial_charge_2013}, the dephasing time is:
\begin{equation}
    \frac{1}{T_2^\ast}=\frac{\sqrt{\langle (\delta E)^2 \rangle}}{\sqrt{2}\hbar}
\end{equation}
where $\delta E$ are the energy fluctuations of each transition. These fluctuations can be written as:

\begin{equation}
    \delta E = \delta \varepsilon \frac{dJ}{d\varepsilon} \frac{dE}{dJ} + \delta E_{\Delta Z} \frac{dE}{d(\delta b_\perp)} +  \delta E_{Z} \frac{dE}{d\bar{b}}
\end{equation}

where $\delta \varepsilon$ are the fluctuations in detuning, $\delta E_{\Delta Z}$ are the fluctuations on $b_\perp$ and  $ \delta E_{Z}$ the fluctuations of the total Zeeman energy.

By squaring $\delta E$ we can express the terms in as  root-mean-square (r.m.s.): 
\begin{equation}
    \langle \delta E^2 \rangle = 
     \delta \varepsilon_{rms}^2  \left(\frac{dJ}{d\varepsilon} \frac{dE}{dJ}\right)^2  + 
     \delta E_{\Delta Z_{rms}}^2 \left(\frac{dE}{d(\delta b_\perp)}\right)^2  +
     \delta E_{Z_{rms}}^2  \left(\frac{dE}{d\bar{b}}\right)^2 
\end{equation}

With the analytical formulas presented in Methods, one can obtain $\langle \delta E^2 \rangle$ for the measured transitions.

\begin{table}[h]
\centering
\begin{tabular}{|c|c|c|c|c|}
\hline
 B field direction & Transition  & $\delta \varepsilon_{rms}$ ($\mu$eV) & $\delta E_{Z_{rms}}$ (neV) & $\delta E_{\Delta Z_{rms}}$ (neV) \\
\hline
In-plane & $T_--S$ & $42.5 \pm 5.5$ & $1.3 \pm 0.2$ & $0.2 \pm 0.2$ \\
\hline
In-plane & $T_--T_0$ & $38.1 \pm 3.5$ & $0.3 \pm 0.2$ & $0.1 \pm 0.5$ \\
\hline
In-plane & $T_--T_+$ & $37.1 \pm 3.7$ & $0.0 \pm 0.1$ & $0.1 \pm 0.6$ \\
\hline
Out-of-plane & $S-T_0$ & $36.0 \pm 2.0$ & - & $3.3 \pm 0.2$ \\
\hline
\end{tabular}

\caption{Summary of $\delta \varepsilon_{rms}$, $\delta E_{Z_{rms}}$, and $\delta E_{\Delta Z_{rms}}$ for the fits in Fig.~\ref{Fig:Fig4}(a) for $T_--S$, $T_--T_0$ and $T_--T_+$, and from Fig.~\ref{Fig:Fig4}(c) for $S-T_0$. It is important to note that $\delta E_{\Delta Z_{rms}}$ and $\delta E_{Z_{rms}}$ are influenced by both magnetic field noise via hyperfine interactions and charge noise via g-factors.}
\label{table:T2_inplane_params}
\end{table}

\newpage
\subsection*{Raw dephasing time data.}
Here we present the measured Ramsey oscillations from which we extract the inhomogeneous dephasing time $T_2^\ast$ by fitting the data to the expression $\text{A} \text{cos}(\omega t ) e^{-t/T_2^\ast} +\text{B} $. In this section we present the measurements and the fits from \ref{Fig:Fig4}.

\begin{figure}[H]
\centering
\includegraphics[width=\linewidth]{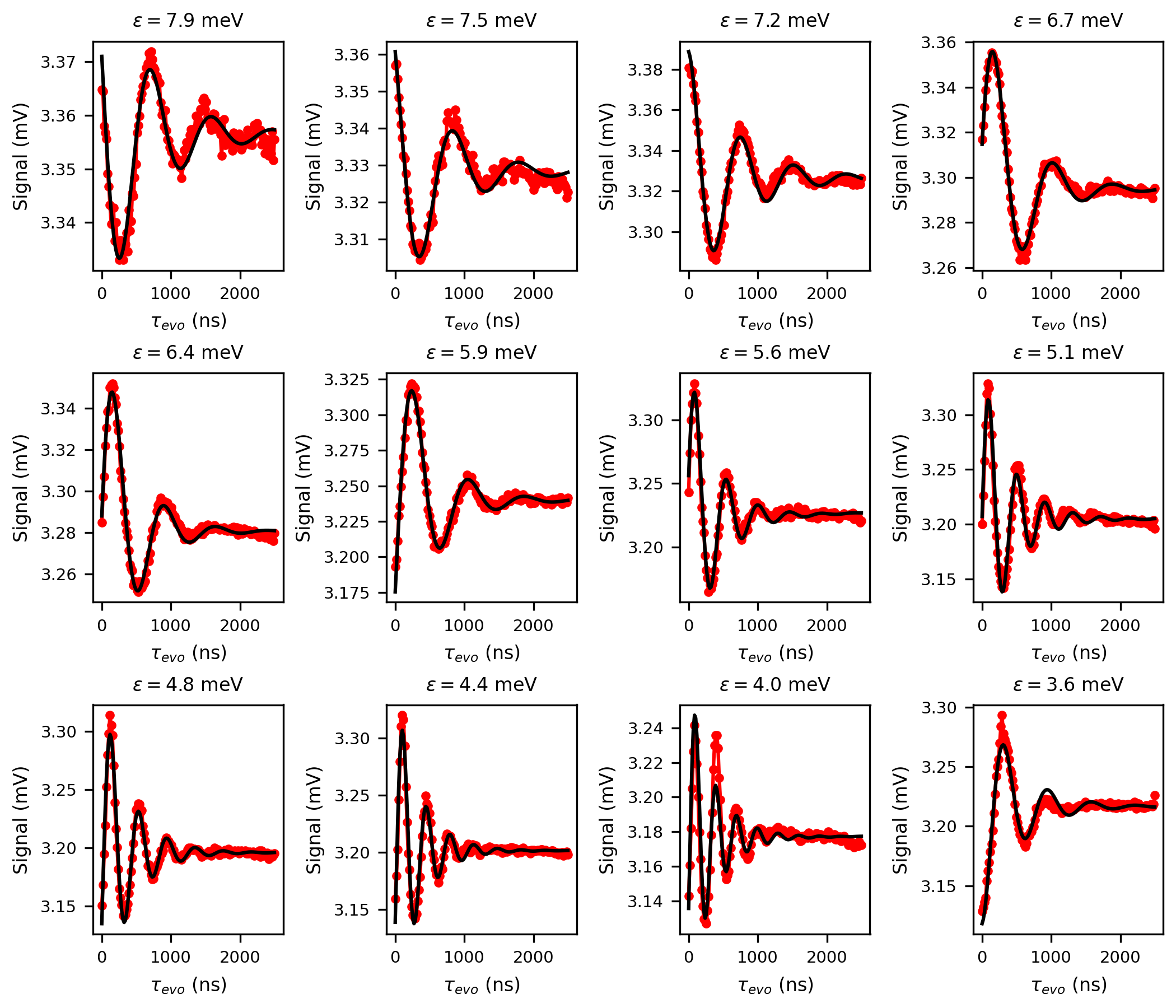}
\caption{Ramsey oscillations at $B_y=30$ mT for the $T_--S$ transition. Each trace is taken at a different detuning point and the total integration time is 20 minutes. Solid lines are the fits to $\text{A} \text{cos}(\omega t ) e^{-t/T_2^\ast} +\text{B} $. Each of these points is represented in Fig.~\ref{Fig:Fig4}(a)}
\label{SFig:SuppT2_T1}
\end{figure}

\begin{figure}[H]
\centering
\includegraphics[width=\linewidth]{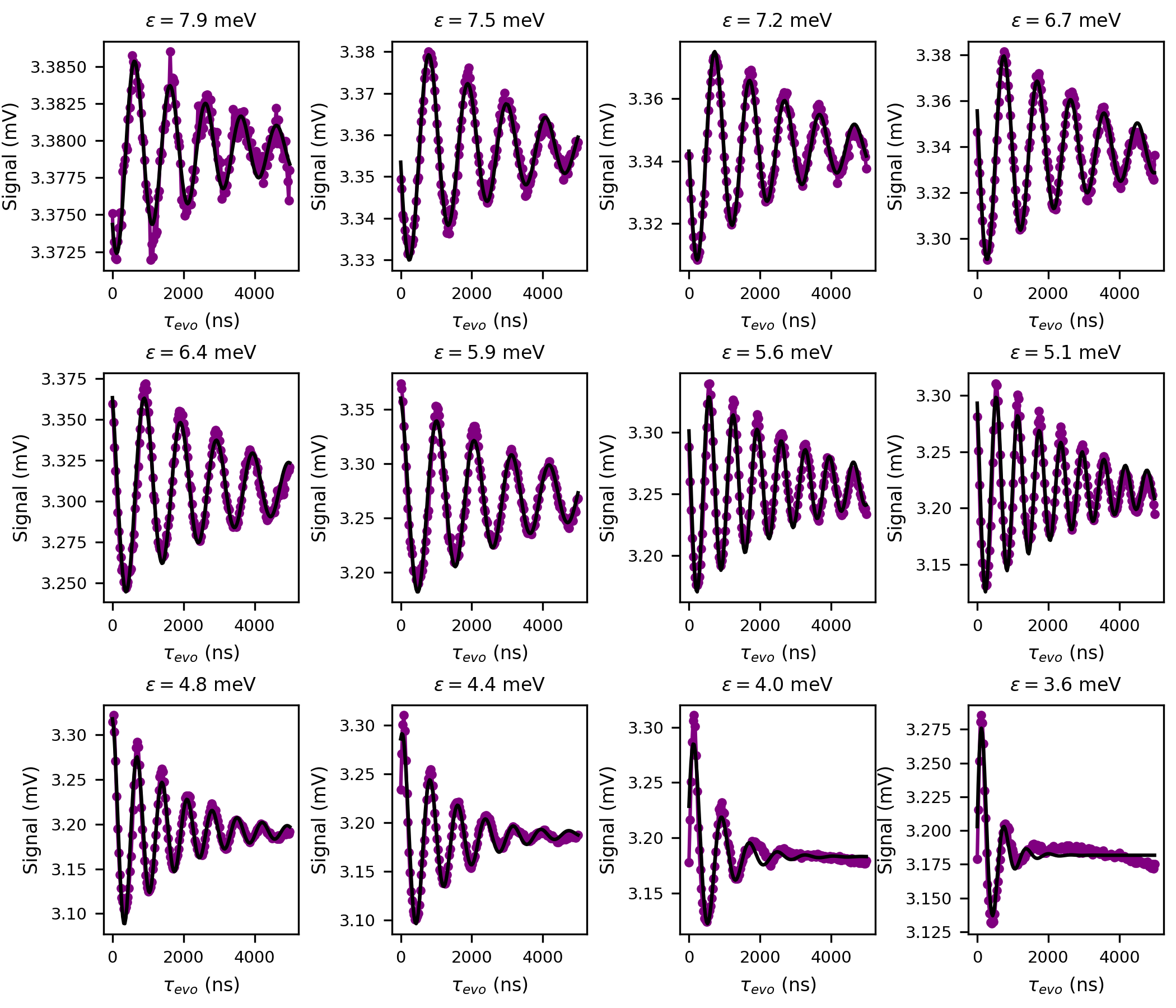}
\caption{Ramsey oscillations at $B_y=30$ mT for the $T_--T_0$ transition. Each trace is taken at a different detuning point and the total integration time is 20 minutes. Solid lines are the fits to $\text{A} \text{cos}(\omega t ) e^{-t/T_2^\ast} +\text{B} $. Each of this points is represented in Fig.~\ref{Fig:Fig4}(a)}
\label{SFig:SuppT2_T2}
\end{figure}

\begin{figure}[H]
\centering
\includegraphics[width=\linewidth]{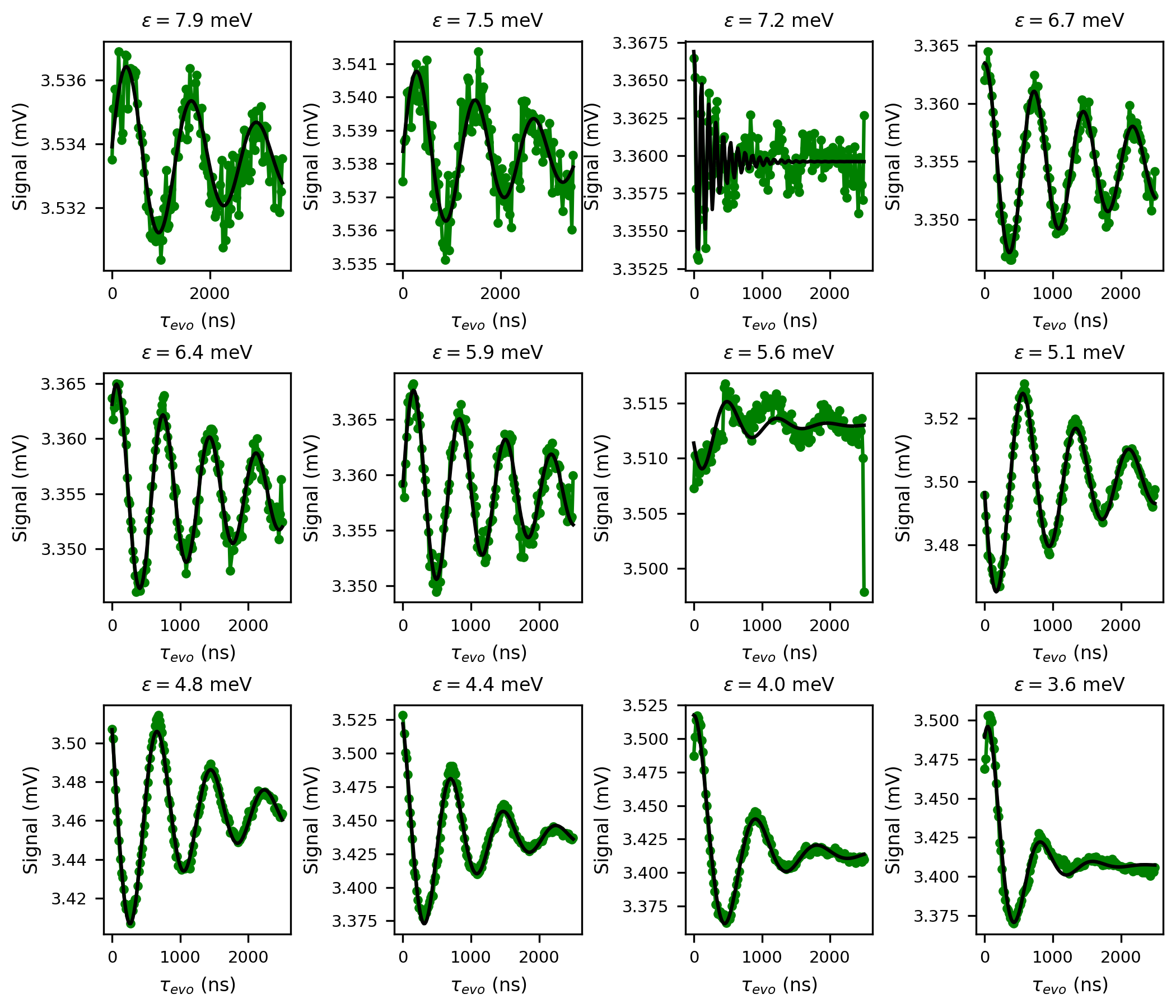}
\caption{Ramsey oscillations at $B_y=30$ mT for the $T_--T_+$ transition. Each trace is taken at a different detuning point and the total integration time is 20 minutes. Solid lines are the fits to $\text{A} \text{cos}(\omega t ) e^{-t/T_2^\ast} +\text{B} $. Each of this points is represented in Fig.~\ref{Fig:Fig4}(a). The panels at $\varepsilon=$7.2 meV and $\varepsilon=$5.6 meV correspond to the outliers in Fig.~\ref{Fig:Fig4}(a).}
\label{SFig:SuppT2_T3}
\end{figure}

\begin{figure}[H]
\centering
\includegraphics[width=\linewidth]{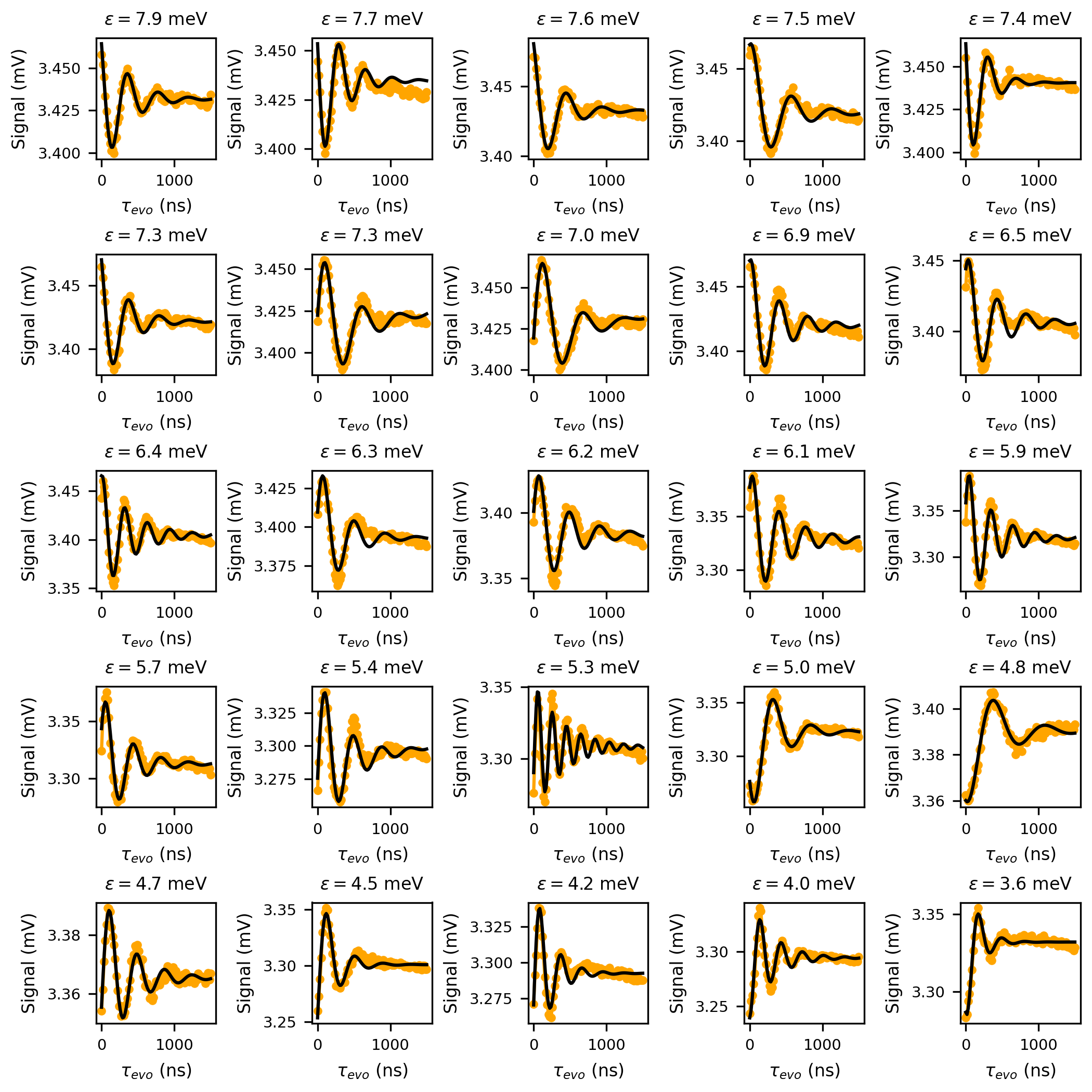}
\caption{Ramsey oscillations at $B_z=12.5$ mT for the $S-T_0$ transition. Solid lines are the fits to $\text{A} \text{cos}(\omega t ) e^{-t/T_2^\ast} +\text{B} $. Each of this points is represented in Fig.~\ref{Fig:Fig4}(c).}
\label{SFig:SuppOUT_fixedB}
\end{figure}

\begin{figure}[H]
\centering
\includegraphics[width=\linewidth]{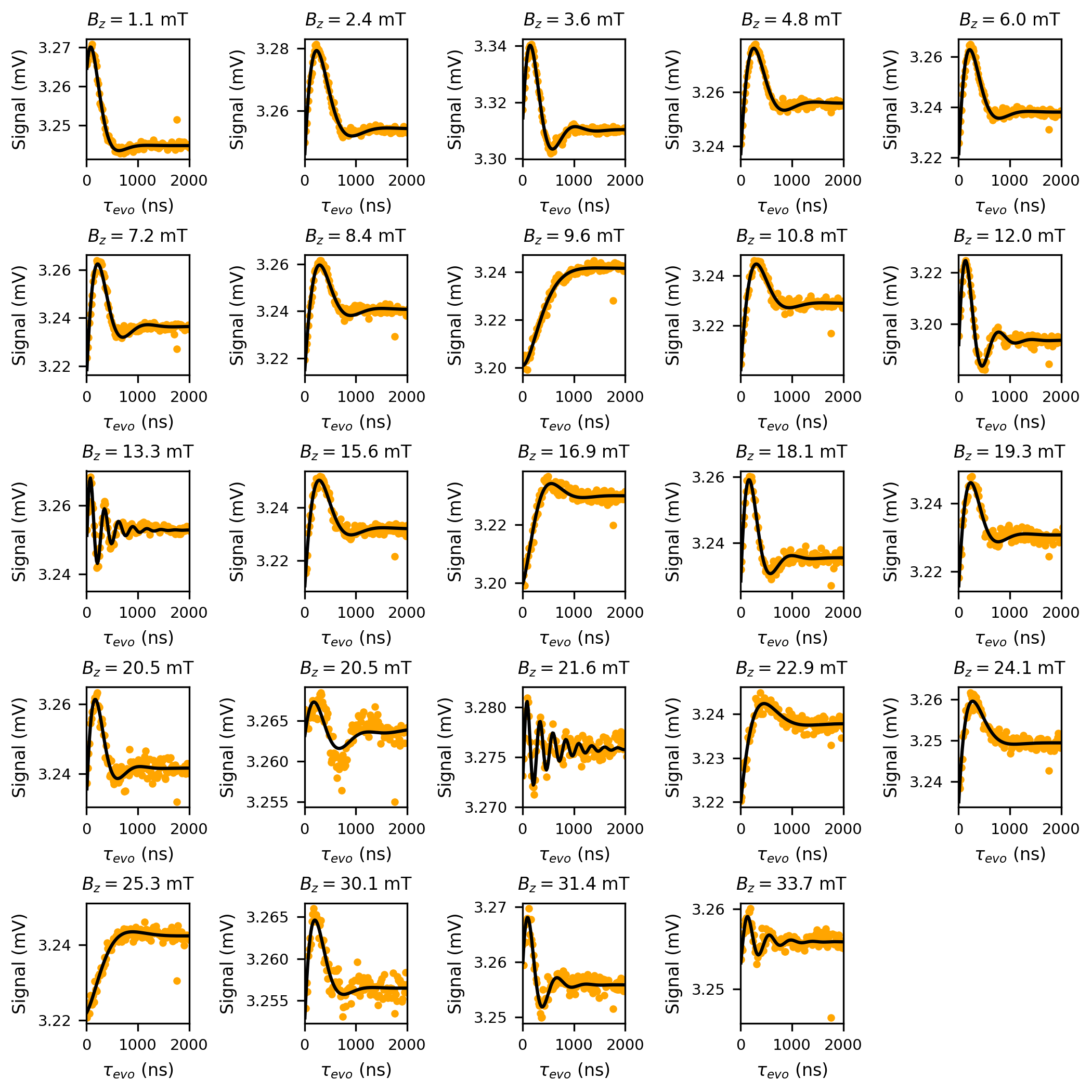}
\caption{Ramsey oscillations at $\varepsilon=7.4$ meV for the $S-T_0$ transition. Solid lines are the fits to $\text{A} \text{cos}(\omega t ) e^{-t/T_2^\ast} +\text{B} $. Each of this points is represented in Fig.~\ref{Fig:Fig4}(d).}
\label{SFig:SuppOUT_fixedeps}
\end{figure}

\end{document}